\documentclass[aps, prb, reprint, superscriptaddress, showpacs]{revtex4-2}

\usepackage{graphicx}                                   
\usepackage{amsmath}                                    
\usepackage{amssymb}                                    
\usepackage[colorlinks=true, allcolors=blue]{hyperref}  
\usepackage{braket}                                     
\usepackage{float}                                      
\usepackage{calc}                                       
\usepackage{stfloats}                                   
\usepackage{orcidlink}                                  
\usepackage[normalem]{ulem}                             
\usepackage{color}

\graphicspath{ {./images/} }
\begin{document}

\newlength{\spinheight}
\settoheight{\spinheight}{\includegraphics[width=0.18\textwidth]{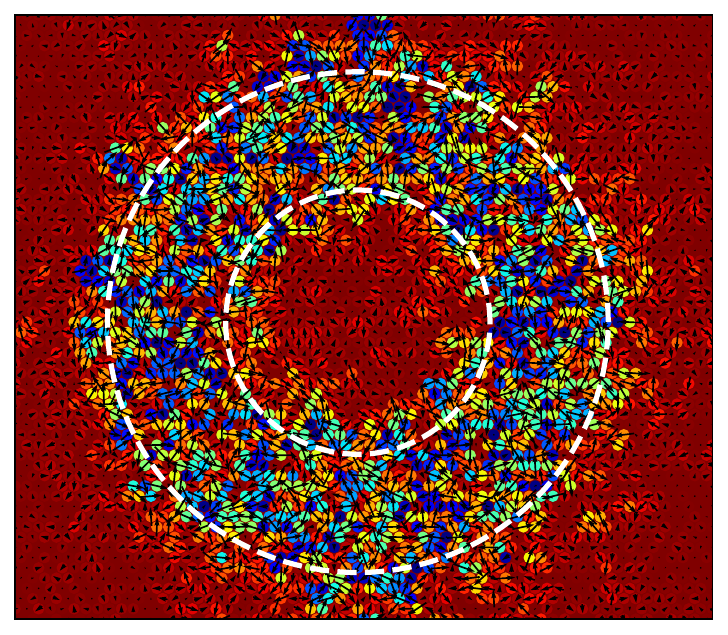}}

\newlength{\probheight}
\settoheight{\probheight}{\includegraphics[width=0.18\textwidth]{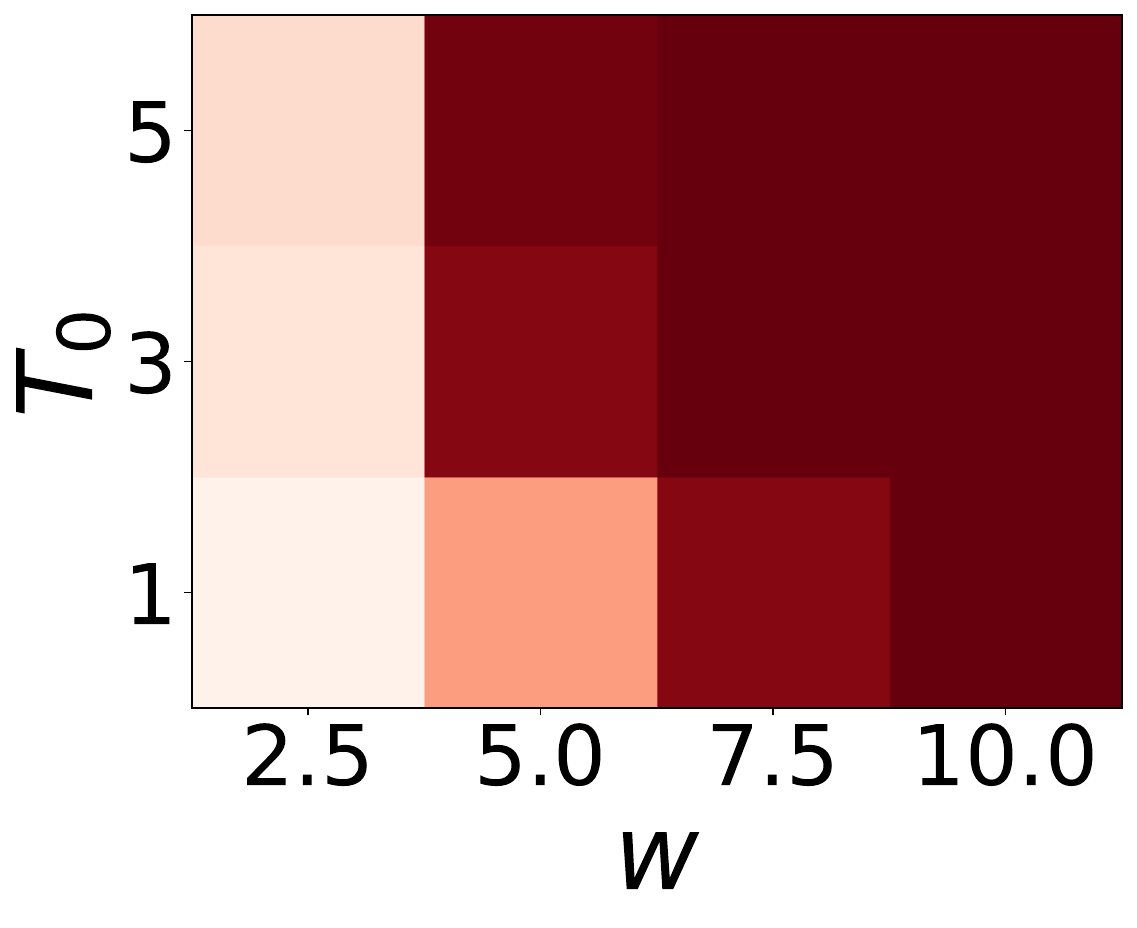}}

\newlength{\jaheight}
\settoheight{\jaheight}{\includegraphics[width=0.18\textwidth]{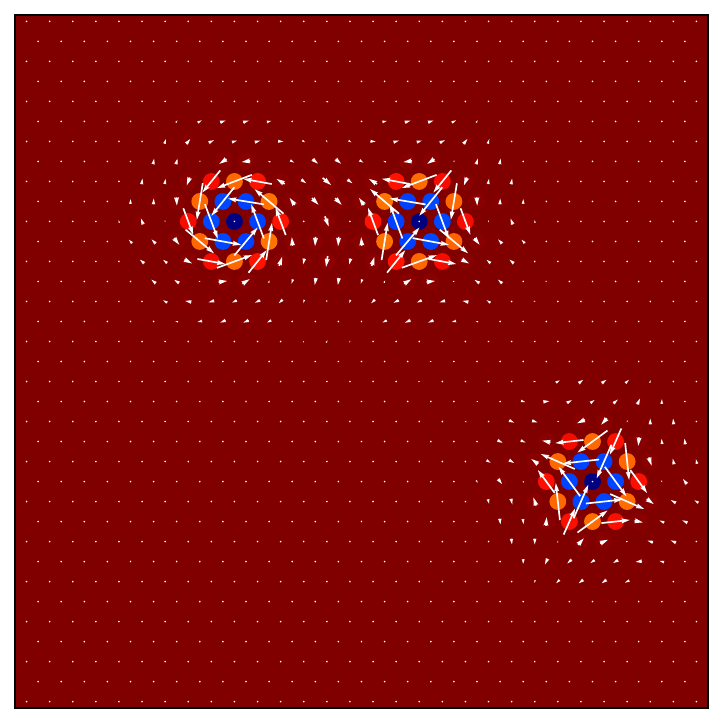}}

\newlength{\beamheight}
\settoheight{\beamheight}{\includegraphics[width=0.2\textwidth]{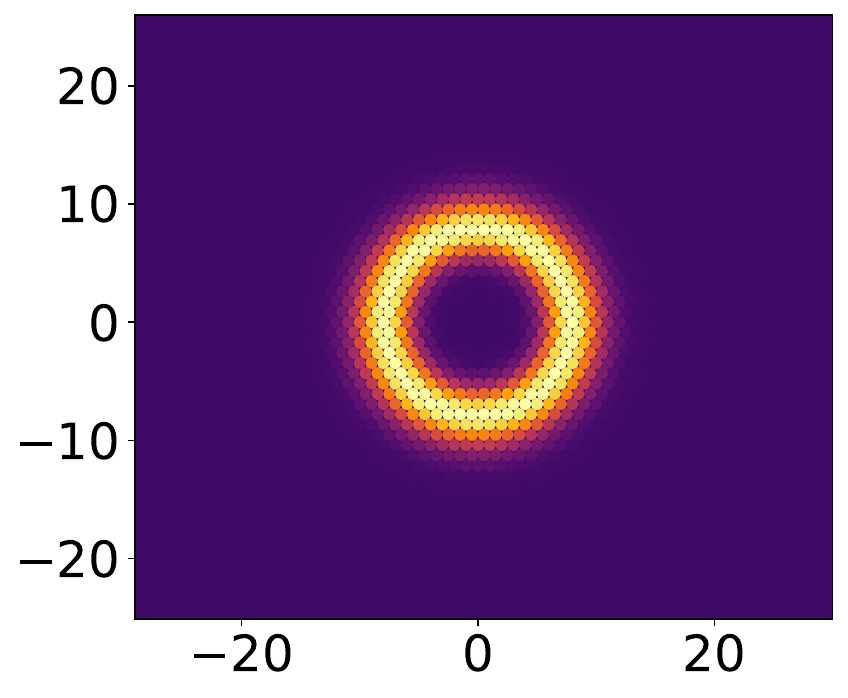}}

\title{Skyrmion generation via Laguerre-Gaussian beam irradiation \\ in frustrated magnets}

\author{Reivienne Jei Laxamana
\orcidlink{0009-0004-8294-0801}}
\email{laxamana@phys.sci.hokudai.ac.jp}

\author{Satoru Hayami
\orcidlink{0000-0001-9186-6958}}
\email{hayami@phys.sci.hokudai.ac.jp}
\affiliation{Department of Physics, Hokkaido University, Sapporo 060-0810, Japan}

\date{\today}

\begin{abstract}
Since its discovery, the study of magnetic skyrmions has been on the rise. In this paper, we discuss our investigations on the light-induced mechanisms for skyrmion generation in a centrosymmetric triangular magnetic lattice with competing $J_1$--$J_3$ interactions, and easy-axis anisotropy. We solve the stochastic Landau-Lifshitz-Gilbert equation for the lattice spin dynamics under Laguerre-Gaussian beam irradiation. Numerical results show that skyrmions are nucleated in two thermodynamic regions, each favoring different phases: the ferromagnetic phase and the skyrmion-lattice phase. 
In the ferromagnetic region, isolated skyrmions are generated mainly through stochastic thermal nucleation. In this regime, higher temperatures and larger beam widths are required to overcome the nucleation barrier. In contrast, in the skyrmion-lattice region, skyrmion nucleation occurs via thermal annealing, where the system relaxes toward its true ground state.
These findings establish a comprehensive theoretical framework for optimizing optical control to generating light-induced skyrmionic textures in frustrated magnets.
\end{abstract}

\maketitle




\section{Introduction}

A magnetic skyrmion is a topological excitation with a noncoplanar vortex geometry~\cite{bogdanov1989, bogdanov1994, rossler2006spontaneous, nagaosa2013}.
Conceptualized in 1961~\cite{bogdanov1989}, they have been discovered within helimagnets in 2009~\cite{muhlbauer2009,yux2010}.
The discovery of skyrmions highlighted the importance of topology in condensed matter systems, where nontrivial spin textures are characterized by quantized topological invariants ~\cite{dirac1931, skyrme1962}.
The spin structure of a skyrmion induces an effective magnetic flux on spin-polarized electrons and magnons, leading to the topological Hall effect~\cite{liy2013, neubauer2009, lee2009, zang2011, schulz2012, hoogdalem2013, mochizuki2014}.
The motion of a skyrmion can be manipulated by electric current, thermal gradient, strain, etc~\cite{jonietz2010, yux2012, schulz2012, kong2013, shizeng2014, mochizuki2014, shibata2015}.
In particular, skyrmions that move in low electric currents have sparked the development of skyrmionics, with the aim of developing next-generation data devices~\cite{iwasaki2013, zhou2014, tomasello2014, fert2013, nagaosa2013, zhang2020skyrmion}.

Most of the earlier research on skyrmions has focused on the role of the Dzyaloshinskii-Moriya (DM) interaction in skyrmion formation~\cite{dmi1958, moriya1960a, moriya1960b}.
Although the DM interaction only directly promotes the spiral spin configuration, a skyrmion lattice can arise if both the DM interaction and an optimal magnetic field are applied to a triangular lattice~\cite{rossler2006spontaneous,Yi_PhysRevB.80.054416,butenko2010, rossler2011}.
Increasing the magnetic field further above a critical field results in the transition to the ferromagnetic state, where isolated skyrmions can exist as topologically protected defects~\cite{butenko2010, rossler2011}.
However, more recent research has shifted focus towards other mechanisms of skyrmion formation in centrosymmetric magnets without the DM interaction~\cite{Okubo_PhysRevLett.108.017206, kurumaji2019skyrmion, hirschberger2019skyrmion, Hirschberger_10.1088/1367-2630/abdef9, hayami2024stabilization}.
Several studies have shown that easy-axis anisotropy in frustrated triangular lattices can stabilize skyrmions in a manner similar to the DM interaction in noncentrosymmetric magnets~\cite{leonov2015, shizeng2016, Hayami_PhysRevB.93.184413, batista2016frustration, Lin_PhysRevLett.120.077202, Hayami_PhysRevB.99.094420, Hayami_PhysRevB.103.224418, kawamura2025frustration}.

Beyond skyrmion generation and stability, a recent topic of interest in skyrmionics is the manipulation of skyrmionic textures through external stimuli such as electric~\cite{Finazzi_PhysRevLett.110.177205,mochizuki2015writing, mochizuki2016creation, je2018creation, Berruto_PhysRevLett.120.117201, Lin_PhysRevResearch.3.023055, Yambe_PhysRevB.110.014428, Furuya_PhysRevResearch.6.013228} and optical fields~\cite{fujita2017}.
In the context of optical fields, Fujita and Sato have demonstrated that structured optical fields provide an effective route to imprint beam profiles onto a chiral magnetic field and thereby generate skyrmionic defects~\cite{fujita2017, Fujita_PhysRevB.96.060407}.
Through irradiation under different modes of the Laguerre--Gaussian (LG) beam, which is characterized by its vanishing intensity at the center, they suceeded in creating conventional skyrmions, skyrmioniums ($2\pi$ vortices), and higher-order skyrmion multiplexes ($n\pi$ vortices) on magnets with strong DM interaction.
Motivated by these results, we investigate whether a similar optical imprinting mechanism can operate in anisotropy-based frustrated magnets, where skyrmions have additional degrees of freedom: vorticity and helicity. 

To address this question, we consider the irradiation of LG beams onto a frustrated magnetic sheet without DM interaction and examine the success of generating topological textures. 
In our simulations, the LG beam irradiation is modeled as a stochastic, spatially non-uniform thermal field~\cite{fujita2017}.
We then investigate the resulting spin dynamics by solving the stochastic Landau--Lifshitz--Gilbert (sLLG) equation.
Our numerical results demonstrate that LG-beam irradiation can indeed induce two distinct types of skyrmionic spin textures depending on the underlying magnetic ground-state phase: isolated skyrmions in the high-field regime and skyrmion lattices in the intermediate-field regime.
These spin textures emerge through different physical mechanisms. 
In the high-field regime, isolated skyrmions are created via stochastic nucleation, where randomly excited local spots evolve and stabilize into topologically nontrivial defects. 
In contrast, the intermediate-field region favors the generation of skyrmion lattice. As such, the LG beam-induced heating acts as an annealer, driving the system from the metastable ferromangetic state towards the true ground state, i.e., the skyrmion lattice.
Furthermore, we find that introducing bond-dependent planar anisotropy in conjunction with LG-beam irradiation significantly suppresses the occurrence of antiskyrmions with the negative vorticity, thereby enhancing the selective generation of skyrmions with the positive vorticity.

The rest of the paper is organized as follows. In Sec.~\ref{sec:methods}, we discuss the data-generation methods used in the study.
Specifically, we explain the triangular Heisenberg model, the LG beam profile, and the variational calculations that allowed us to attain and interpret our results.
Sections~\ref{sec:isolated} and~\ref{sec:skx} dive deeper into the generation mechanism of two skyrmionic spin textures: isolated skyrmion and skyrmion lattice.
Section~\ref{sec:ja} discusses the influence of bond-dependent planar anisotropy on the distinction of skyrmions and antiskyrmions.
Finally, a conclusion was reached in Sec.~\ref{sec:conclusion}, where we summarize our findings.

\section{Methods} \label{sec:methods}

This section explains the methodologies used in the research. Section~\ref{sec:model} describes the Hamiltonian model used in the simulations.
Section~\ref{sec:beam} discusses the LG beam and it's properties.
Finally, the phase diagram attained through variational calculations can be found in Sec.~\ref{sec:var_calc} along with the ansatz used in the calculations.

\subsection{Model} \label{sec:model}

We use a frustrated triangular-lattice model with easy-axis anisotropy.
\begin{equation}
    \mathcal{H} = \sum_{\braket{i, j}} J_{ij} \mathbf{S}_i\cdot\mathbf{S}_j - H\sum_i S_i^z - A\sum_i (S_i^z)^2.
    \label{eq:model}
\end{equation}
Each spin vector $\mathbf{S}_i$ is normalized with a magnitude of $1$.
The first term denotes the isotropic exchange interaction which includes the nearest-neighbor ferromagnetic coupling $J_1<0$ and third-nearest-neighbor antiferromagnetic coupling $J_3>0$.
We set $J_1=-1$ to be the unit of energy and the lattice constant $a$ is the unit of length in the simulations.
The second term denotes the Zeeman coupling due to an external magnetic field $H$, while the third term denotes the easy-axis anisotropy with anisotropic constant $A > 0$.

Due to the $C_6$ symmetry of the triangular lattice, there are six possible ordering wave vectors $\mathbf{q}_v = \pm Q(\cos 2u\pi/3, \sin 2u\pi/3)$ where $Q= 2\cos^{-1}[(1+\sqrt{1-2J_1/J_3})/2]$ and $v=u+1=1, 2, 3$.
These wave vectors are parallel to the nearest-neighbor bond directions.
To maintain the boundary conditions of a lattice with $N = L^2$ sites, the wave number must be in the form $Q = 2n\pi/L$ with $n=1,2,\dots, L$, where $L$ is the length of one side of the triangular lattice.
Preliminary simulations on various lattice sizes and $J_3/|J_1|$ ratios showed that the conclusive results were independent of these parameters.
Consequently, the lattice size $N$ and $J_3/|J_1|$ ratio was fixed for the findings presented in this paper.
We used a lattice with $N = 60^2 = 3600$ sites resulting in a commensurate ordering wave number $Q=2\pi/5 \; (n=12)$ for $J_3/|J_1| = 0.5$.

\subsection{Beam Profile} \label{sec:beam}

\begin{figure}[H]
    \centering
    \begin{minipage}{0.2\textwidth}
        \centering
        $p=0, m=5, w=5$\\
        \includegraphics[width=\linewidth]{profiles/p0_m5_w5.pdf}
    \end{minipage}
    \begin{minipage}{0.2\textwidth}
        \centering
        $p=0, m=5, w=7.5$\\
        \includegraphics[width=\linewidth]{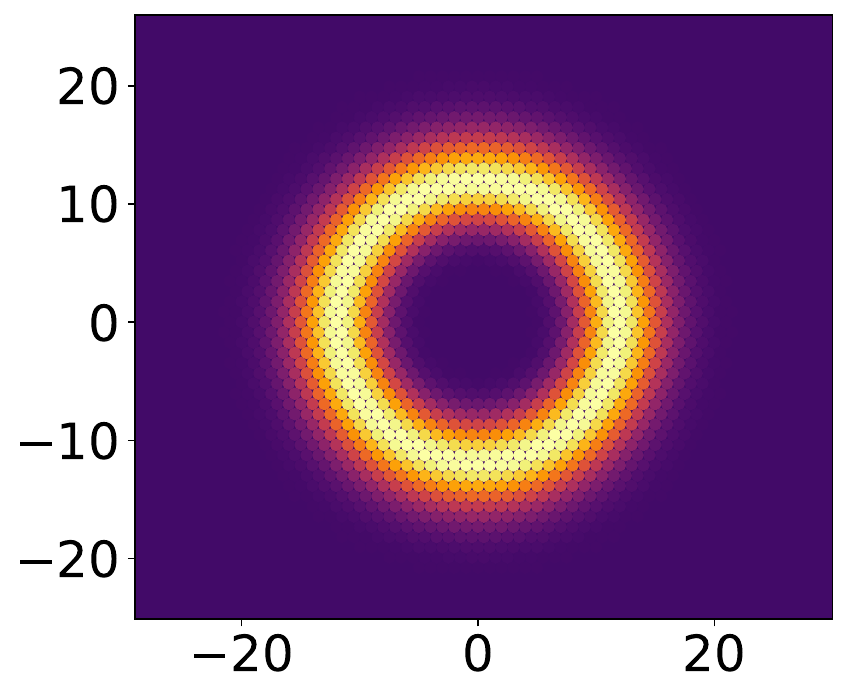}
    \end{minipage}
    \begin{minipage}{0.2\textwidth}
        \centering
        $p=1, m=5, w=7.5$\\
        \includegraphics[width=\linewidth]{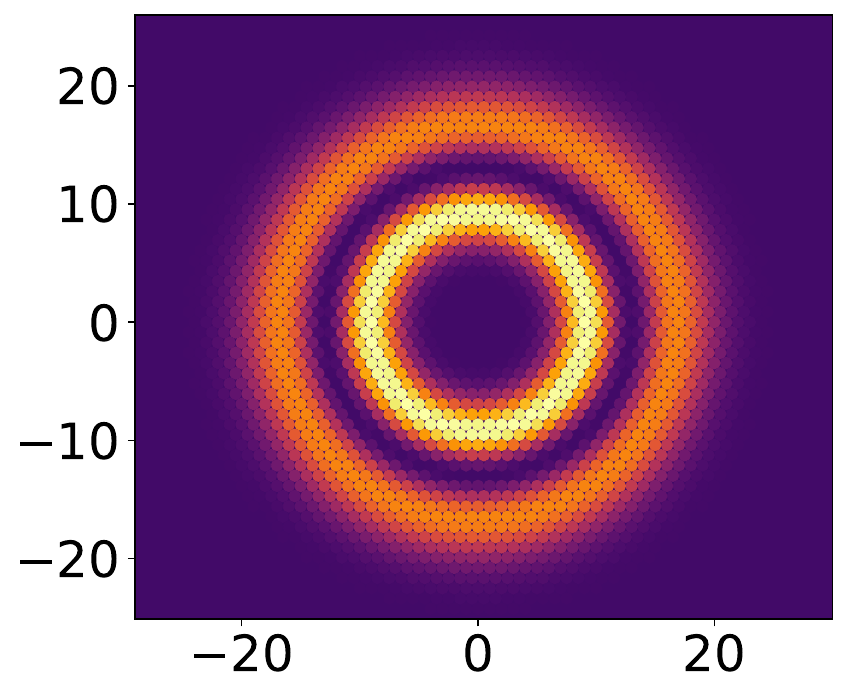}
    \end{minipage}
    \begin{minipage}{0.2\textwidth}
        \centering
        $p=2, m=5, w=7.5$\\
        \includegraphics[width=\linewidth]{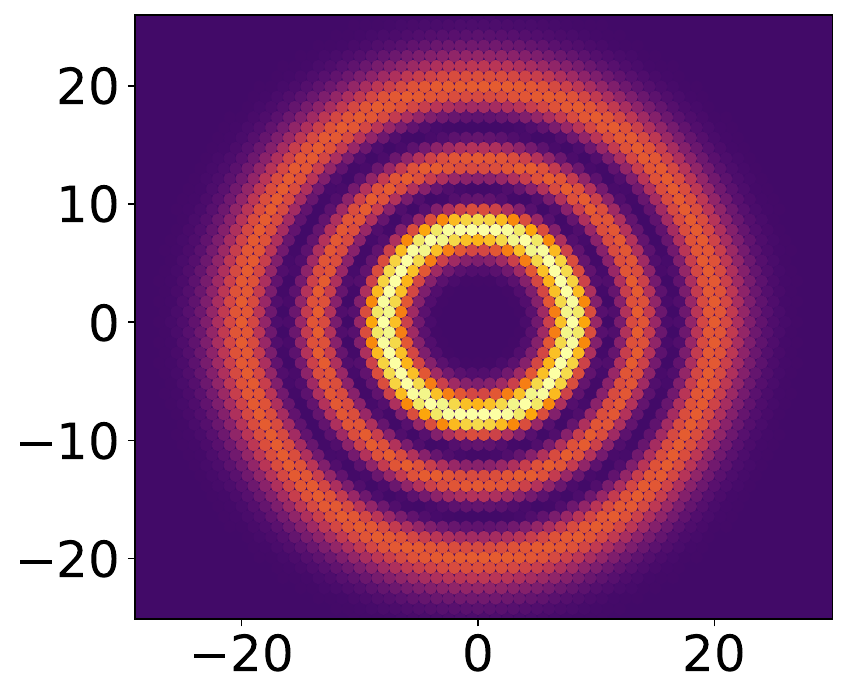}
    \end{minipage}
    \begin{minipage}{0.1\textwidth}
        \vspace{-5pt}
        \includegraphics[height=1.15\beamheight]{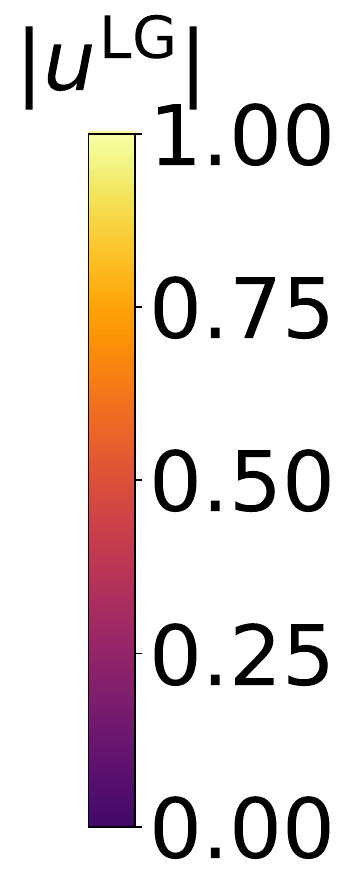}
    \end{minipage}
    \caption{Beam profiles at different $(p, m, w)$. The colors denote the local intensity at that site in the triangular lattice.}
    \label{fig:beam_profiles}
\end{figure}

The LG beams carry intrinsic orbital angular momentum associated with the azimuthal phase factor, $e^{im\phi}$, with azimuthal quantum number $m$~\cite{wang2018recent, Quinteiro_RevModPhys.94.035003}.
For $m\neq 0$, this phase structure leads to a phase singularity on the beam axis ($\rho=0$).
To keep the electromagnetic field finite and single-valued, the amplitude necessarily vanishes at the origin, resulting in a characteristic doughnut-shaped intensity profile.
We used the following LG beam equation in cylindrical coordinates as
\begin{equation}
    u^\text{LG} (\rho, \phi) = \frac{1}{\sqrt{|w|}}\left(\frac{\rho}{w}\right)^{|m|} e^{-\frac{2\rho^2}{w^2}+im\phi} \left[L_p^{|m|}\left(\frac{2\rho^2}{w^2}\right)\right],
    \label{eq:lg-beam}
\end{equation}
where $L^{|m|}_p$ is the generalized Laguerre function and $p$ is the radial mode index.
The LG equation is evaluated at $z=0$, as it is where the beam waist $w$ is at its minimum($w=w_0$). 
Away from $z=0$, the beam radius expands according to $w(z) = w_0\sqrt{1+|z|/Z}$, where $Z$ is the Rayleigh range, defined as the propagation distance at which the cross-sectional area of the beam increases by a factor of two, i.e., from $\pi w_0^2$ to $2\pi w_0^2$. 
For visualization, Fig.~\ref{fig:beam_profiles} displays the intensity profiles of LG beams for several representative modes characterized by different values of $p$ and $w$.
In the present study, the effect of LG-beam irradiation is incorporated as a thermal perturbation through a spatially and temporally dependent temperature function $T(\mathbf{r},t)$~\cite{fujita2017},
\begin{equation}
    T(\mathbf{r}, t) = u^\text{LG} (\rho, \phi) T_0 \left(1 - \frac{t}{t_0}\right) \theta(t_0-t)\theta(t),
\end{equation}
where $\theta(t)$ denotes the Heaviside step function, $T_0$ is the beam temperature, and $t_0$ is the duration of beam irradiation.
The thermal function can be used in the fluctuation-dissipation theorem to attain the thermal profile $\mathbf{h}_T(t)$, which is given by
\begin{equation}
    \mathbf{h}_T(t) = \frac{2k_{\mathrm{B}} 
    T(\mathbf{r}, t)\alpha}{\hbar\gamma^2},
\end{equation}
where $k_{\mathrm{B}}
$ is the Boltzmann constant, $\alpha$ is the Gilbert damping constant, and $\gamma$ is the gyromagnetic ratio.

The Hamiltonian in Eq.~\eqref{eq:model} and the thermal profile is used in the sLLG equation, 
\begin{equation}
    \frac{d\mathbf{S}}{dt} = -\gamma\mathbf{S}\times\left(-\frac{\partial \mathcal{H}}{\partial\mathbf{S}} + \mathbf{h}_T(t)\right) + \alpha\mathbf{S}\times\frac{d\mathbf{S}}{dt},
    \label{eq:sllg}
\end{equation}
where Heun's method is applied to give the spin dynamics.
The sLLG equation and the simulations assume natural units $\hbar=1$, $k_{\mathrm{B}}=1$, and $\gamma=1$. Furthermore, our simulations use a fixed $\alpha=0.1$, allowing convergence to a stable state. This is also consistent with other researches using a similar model~\cite{shizeng2016, Liang_2018}.

Each simulation can be divided into two processes: irradiation (heating) and annealing.
Based on initial simulations with various irradiation and annealing times, the irradation time $t_0 = 500$ and annealing time $t_1 = 500$ were found to be optimal.
Succeeding simulations were then done at $t_0 = 500$ and $t_1=500$ for a total simulation time $t_{\rm sim} = t_0+t_1 = 1000$.
The conversion factor $\hbar/J$ is applied to the simulation time to get the real time $t_\text{real} = t_\text{sim} (\hbar/J)$.
At $J = 1$ meV, the simulation time unit would correspond to $t_\text{real} \approx 6.58$ ps. 

\onecolumngrid

\twocolumngrid

\subsection{Variational Calculations} \label{sec:var_calc}

\begin{figure}[H]
    \centering
    \includegraphics[width=0.49\textwidth]{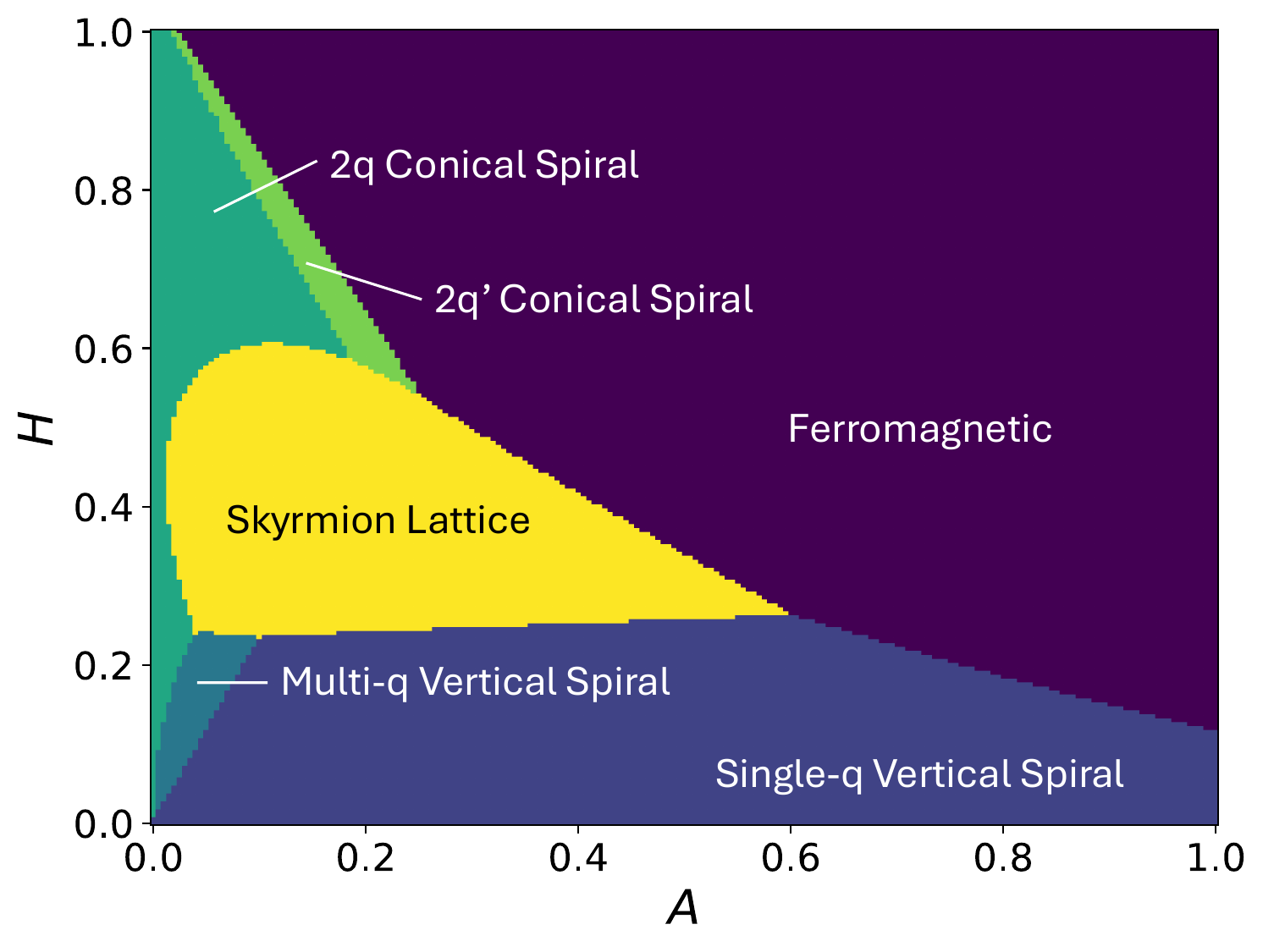}
    \caption{Phase diagram at $T=0$, obtained through variational calculations~\cite{leonov2015, shizeng2016}. 
    A total of $201\times 201$ data points were generated, each comparing the energies of $6$ phases.}
    \label{fig:phase_diagram}
\end{figure}

Before presenting the results under LG-beam irradiation, we first discuss the ground-state phase diagram of the model Hamiltonian in Eq.~\eqref{eq:model} in the absence of the beam~\cite{leonov2015, shizeng2016}.
Figure~\ref{fig:phase_diagram} presents the phase diagram obtained by performing the variational calculations at $T=0$ with $L=60$ ($N = 3600$) spins and commensurate ordering wave number $Q = 2\pi/5$.

The phase diagram was attained through fixed-$q$ variational calculations.
By fixing the wave number to $Q=2\pi/5$ for $J_3/|J_1|=0.5$, the resulting ground-state phases become commensurate with the underlying lattice.
We consider a total of 6 phases~\cite{leonov2015, shizeng2016}, with each spin configuration defined by $\mathbf{S} = \mathbf{M}/|\mathbf{M}|$.

\begin{figure}[H]
    \centering
    \includegraphics[width=0.45\textwidth]{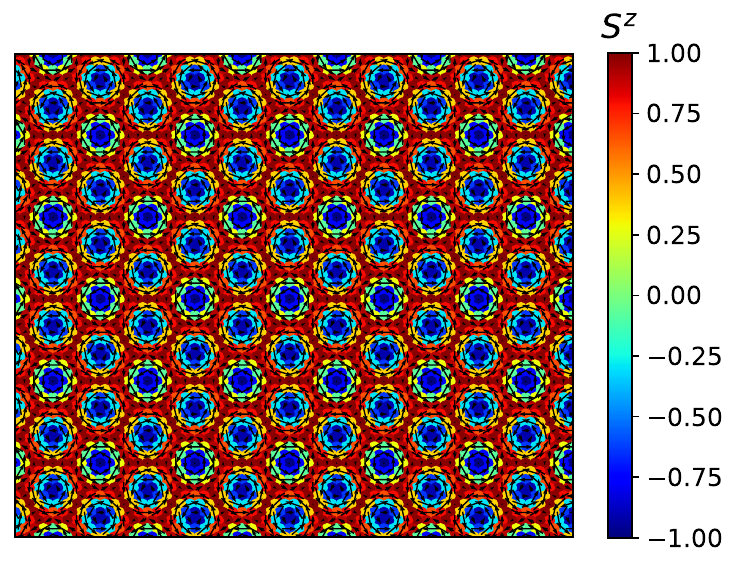}
    \caption{An example of a skyrmion lattice from variational calculations at $a_1 = a_2 = 0.8$ and $\bar{m} = 0.2$.}
    \label{fig:lattice_var}
\end{figure}

\begin{enumerate}
    \item {
        Ferromagnetic:
        \begin{equation}
            M_i^{z} = 1,
            \quad
            M_i^{x} = M_i^{y} = 0
        \end{equation}
    }
    \item {
        Single-$q$ vertical spiral: the variational parameters are $a_1$, $a_2$, and $\bar{m}$.
        \begin{equation}
            \begin{split}
            &M_i^x = a_1 \sin(\mathbf{q_1}\cdot \mathbf{r}_i) \\
            &M_i^y = 0 \\
            &M_i^z = a_2 \cos(\mathbf{q_1}\cdot \mathbf{r}_i) + \bar{m} 
            \end{split}
        \end{equation}
    }
    \item {
        Multiple-$q$ vertical spiral: the variational parameters are $a_1$, $a_2$, $\phi$ 
        and $\bar{m}$.
        \begin{equation}
            \begin{split}
            &M_i^x = a_1 \cos(\mathbf{q}_2\cdot\mathbf{r}_i + \phi 
            ) - a_1 \cos(\mathbf{q}_3\cdot\mathbf{r}_i - \phi 
            )\\
            &M_i^y = -a_2 \sin(\mathbf{q}_1\cdot\mathbf{r}_i) \\
            &M_i^z = a_2 \cos(\mathbf{q}_1\cdot\mathbf{r}_i) + \bar{m}  
            \end{split}
        \end{equation}
    }
    \item{
        $2q$ conical spiral: a multiple-$q$ spiral configuration wherein $a_1\ne a_2$. The variational parameters are $a_1$, $a_2$, $a_3$, and $\bar{m}$.
        \begin{equation}
            \begin{split}
            &M_i^x = a_1 \cos(\mathbf{q}_1\cdot\mathbf{r}_i) + a_2 \cos(\mathbf{q}_2\cdot\mathbf{r}_i)\\
            &M_i^y = -a_1 \sin(\mathbf{q}_1\cdot\mathbf{r}_i) + a_2 \sin(\mathbf{q}_2\cdot\mathbf{r}_i)\\
            &M_i^z = a_3 \cos(\mathbf{q}_3\cdot\mathbf{r}_i) + \bar{m}  
            \end{split}
        \end{equation}
    }
    \item{
        $2q'$ conical spiral: a multiple-$q$ spiral configuration wherein $a_1=a_2$. The variational parameters are $a_1$, $a_3$, and $\bar{m}$.
        \begin{equation}
            \begin{split}
            &M_i^x = a_1 \cos(\mathbf{q}_1\cdot\mathbf{r}_i) + a_1 \cos(\mathbf{q}_2\cdot\mathbf{r}_i)\\
            &M_i^y = -a_1 \sin(\mathbf{q}_1\cdot\mathbf{r}_i) + a_1 \sin(\mathbf{q}_2\cdot\mathbf{r}_i)\\
            &M_i^z = a_3 \cos(\mathbf{q}_3\cdot\mathbf{r}_i) + \bar{m}  
            \end{split}
        \end{equation}
    }
    \item{
        Skyrmion lattice: the variational parameters are $a_1$, $a_2$, and $\bar{m}$.
        \begin{equation}
            \begin{split}
            &M_i^x = a_1 \sum_{v=1}^3 \left[\sin(\mathbf{q}_v\cdot\mathbf{r}_i)\right] \mathbf{e}_v\cdot\hat{\mathbf{x}} \\
            &M_i^y = a_1 \sum_{v=1}^3 \left[\sin(\mathbf{q}_v\cdot\mathbf{r}_i)\right] \mathbf{e}_v\cdot\hat{\mathbf{y}} \\
            &M_i^z = a_2 \sum_{v=1}^3 \left[\cos(\mathbf{q}_v\cdot\mathbf{r}_i)\right] \mathbf{e}_v\cdot\hat{\mathbf{z}} + \bar{m}
            \end{split}
            \label{eq:skx}
        \end{equation}
        where $\mathbf{e}_1 = \hat{\mathbf{y}} - \hat{\mathbf{z}}$,
        $\mathbf{e}_2 = -\sqrt{3}/2\hat{\mathbf{x}} - 1/2\hat{\mathbf{y}} - \hat{\mathbf{z}}$, and 
        $\mathbf{e}_2 = \sqrt{3}/2\hat{\mathbf{x}} - 1/2\hat{\mathbf{y}} - \hat{\mathbf{z}}$.
        Figure~\ref{fig:lattice_var} shows an example of a skyrmion lattice configuration from this ansatz.
    }
\end{enumerate}

\section{Isolated Skrymions} \label{sec:isolated}

\begin{figure}[H]
    \centering
    \begin{minipage}{0.4\textwidth}
        \centering
        \begin{minipage}{0.45\linewidth}
            \centering
            $t=50$\\
            \includegraphics[width=\linewidth]{time_evol/is_50.pdf}
        \end{minipage}
        \begin{minipage}{0.45\linewidth}
            \centering
            $t=400$\\
            \includegraphics[width=\linewidth]{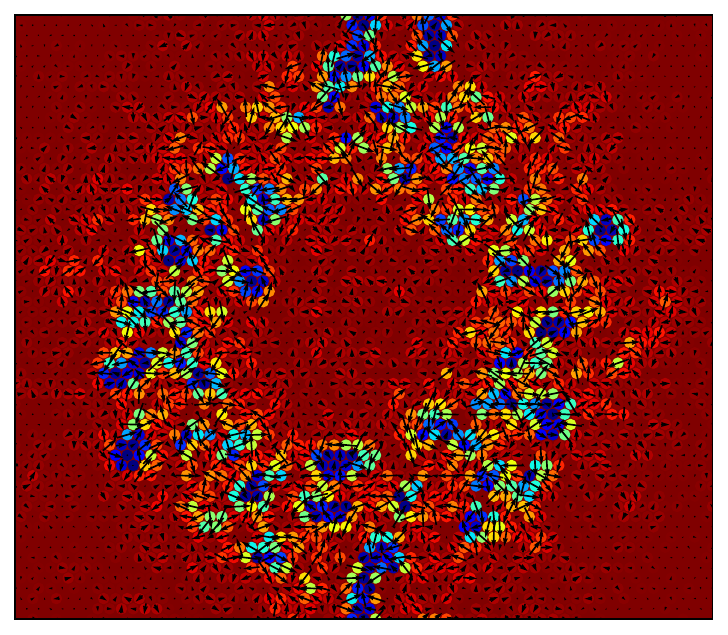}
        \end{minipage} \\
        \begin{minipage}{0.45\linewidth}
            \centering
            $t=500$\\
            \includegraphics[width=\linewidth]{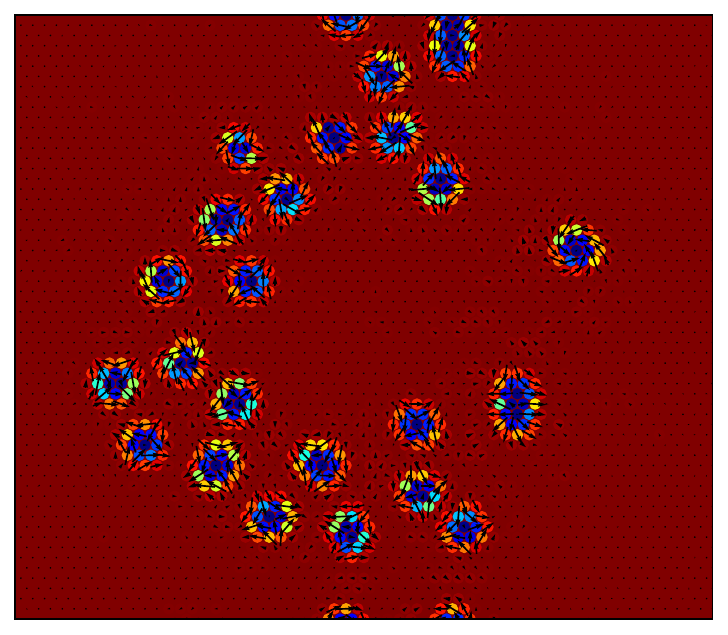}
        \end{minipage}
        \begin{minipage}{0.45\linewidth}
            \centering
            $t=1000$\\
            \includegraphics[width=\linewidth]{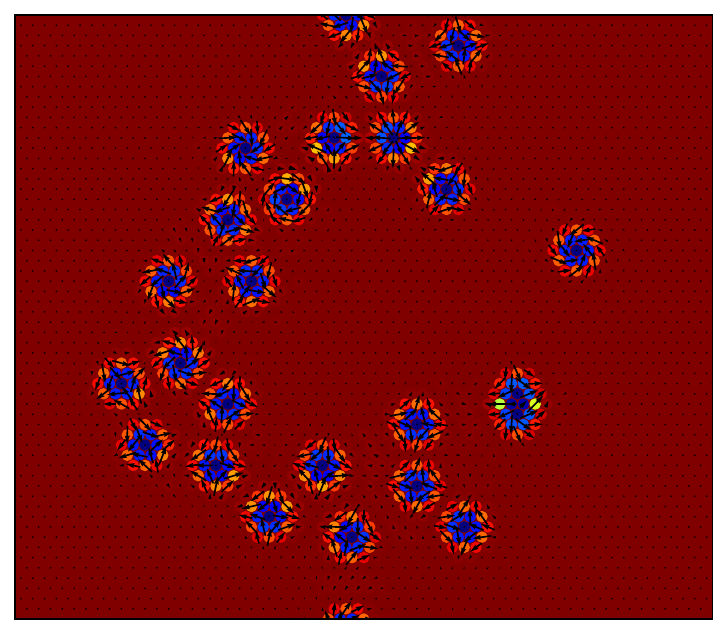}
        \end{minipage}
    \end{minipage}
    \begin{minipage}{0.075\textwidth}
        \includegraphics[height=2.2\spinheight]{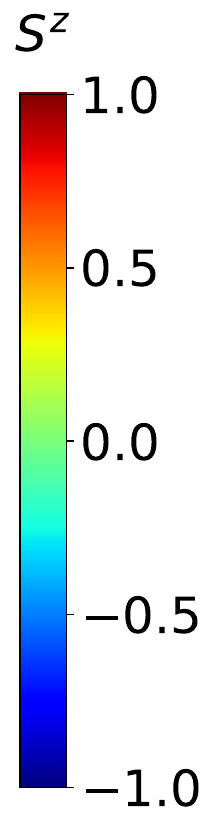}
    \end{minipage}
    \caption{Time evolution of an isolated skyrmion at $A=0.7$ and $H=0.3$ with beam parameters $T_0=5$, $p=0$, $m=5$, and $w=10$. The white dashed circles at $t=50$ represent the upper and lower boundaries of the applied beam. The colorbar represents the out-of-plane spin component $S^z$. 
    }
    \label{fig:is_evol}
\end{figure}

In this section, we discuss the generation of isolated skyrmions.
Figure~\ref{fig:is_evol} illustrates the time evolution of an isolated skyrmion generated by LG beam irradiation in the ferromagnetic regime. 
The simulation is performed at $A=0.7$ and $H=0.3$, where the ferromagnetic state is the ground state in the absence of external perturbations. 
The beam parameters are chosen as $T_0=5$, $p=0$, $m=5$, and $w=10$.
Throughout this work, we set $p = 0$ and $m = 5$, since different choices of these parameters do not qualitatively alter the results.
Before beam irradiation, the initial spin configuration is the ferromagnetic spin state.

At an early stage of irradiation ($t=50$), the LG beam already induces a local reversal of the out-of-plane spin component $S^z$ within the beam-irradiated region around $w(\sqrt{m}/2) \leq \rho \leq w(\sqrt{m}/2 + 1)$.  
This initial spin-flip event serves as the seed for the subsequent formation of the skyrmion core. 
As the heating effect of the LG beam further develops ($t=400$), additional local fluctuations appear within the annular region of the beam profile. 
Subsequently, around the end of the irradiation process ($t=500$), vortex spin textures begin to nucleate, signaling the emergence of skyrmionic defects. 
At this point, the skyrmion boundaries still contain mid-range out-of-plane spin components ($S^z\approx 0$), as represented by the color scale. 
This is in contrast to after the annealing stage ($t=1000$), where the system relaxes into a stable isolated skyrmion embedded in the ferromagnetic background and the out-of-plane spins of the skyrmion boundaries ($S^z \approx 0.8$) and skyrmion core ($S^z \approx -1$) are at extremes.

This extremity in the out-of-plane spins is not found in the skyrmion lattice configurations generated through Eq.~(\ref{eq:skx}) as shown in Fig.~\ref{fig:lattice_var}.
This signals the viability of isolated skyrmion application to next-generation data devices as the extremity allows for a more accurate skyrmion detection via the toplogical Hall effect ~\cite{ludbrook2017nucleation}.
We attribute this extremity in isolated skyrmions to easy-axis anisotropy, as the two regimes dicussed in this section and in Sec.~\ref{sec:skx} differ mainly in easy-axis anisotropy.
High easy-axis anistropy pushes the out-of-plane spins on the skyrmion boundary to align with either extremes, as seen from the $(S^z)^2$ factor in Eq.~\eqref{eq:model}. 
As the high-field regime discussed in this section favors the ferromagnetic spin state, the mid-range out-of-plane spin components are pushed towards one end ($S^z \to 1$).

Overall, these results demonstrate that, even in the absence of the DM interaction, stochastic thermal excitation induced by the LG beam can overcome the nucleation barrier and lead to the formation of isolated skyrmions in frustrated magnets.

\begin{figure}[H]
    \centering
    \begin{minipage}{0.4\textwidth}
        \centering
        \begin{minipage}{0.45\linewidth}
            \centering
            $A=0.7, H=0.3$\\
            \includegraphics[width=\linewidth]{isolated/A=0.7/H=0.3/probability.pdf}
        \end{minipage}
        \begin{minipage}{0.45\linewidth}
            \centering
            $A=0.7, H=0.4$\\
            \includegraphics[width=\linewidth]{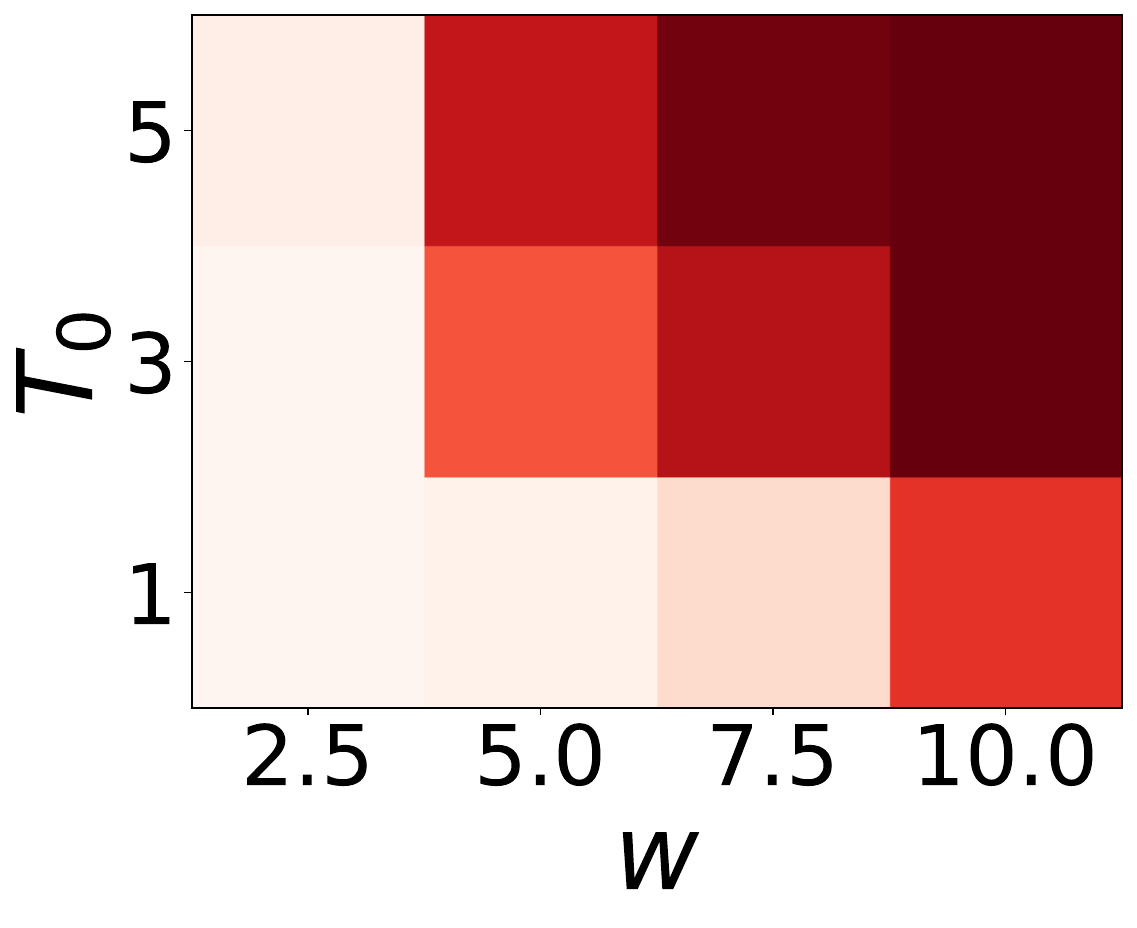}
        \end{minipage}
        \begin{minipage}{0.45\linewidth}
            \centering
            $A=0.7, H=0.5$\\
            \includegraphics[width=\linewidth]{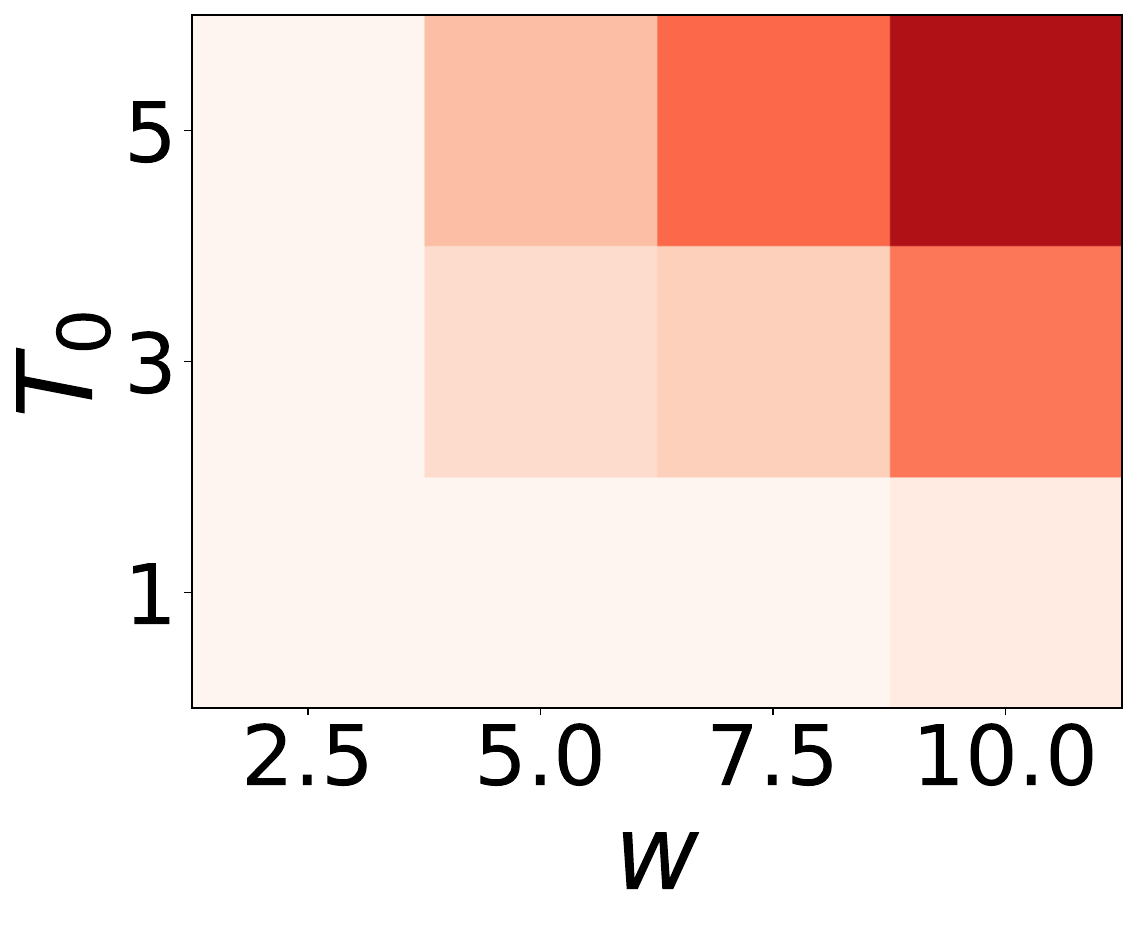}
        \end{minipage}
    \end{minipage}
    \begin{minipage}{0.075\textwidth}
        \vspace{-0.4cm}
        \includegraphics[height=2\probheight]{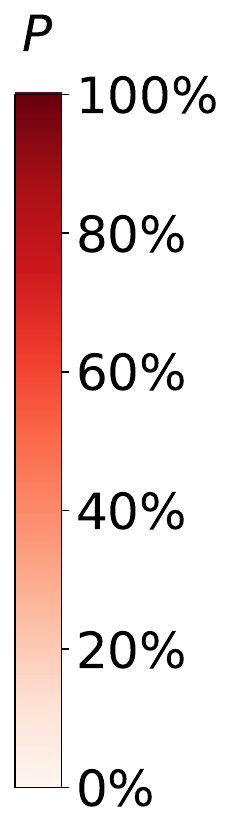}
    \end{minipage}
    \caption{Probability maps at $A=0.7$ and $H=0.3, 0.4$, and $0.5$. The temperature is in units $J_1/k_{\mathrm{B}}$, while the beam width $w$ is in lattice unit $a$.}
    \label{fig:is_H}
\end{figure}

To identify the optimal parameters for skyrmion generation, we show the skyrmion creation probability as a heatmap over various beam parameters in Fig.~\ref{fig:is_H}. 
The skyrmion generation probability $P$ was calculated as the number of successful trials $s$ over the total number of trials $n_{\rm trials} = 50$, converted into percentage,
\begin{equation}
    P = \frac{s}{50}\times100\%.
    \label{eq:prob}
\end{equation}
Each trial operates on a given parameter set but with randomly-varying noise for the thermal field.
To simulate random variation between trials, a seed to the random generator was set for each Heun step $dt$ in each simulation trial.
A successful trial is defined by a nonzero topological charge on the entire lattice, also known as the skyrmion number $n_{\rm sk}$~\cite{BERG1981412},
\begin{equation}
    \begin{gathered}
        n_{\rm sk} = \frac{1}{4\pi}\sum_{\braket{i, j, k}} \arctan \Omega_{ijk}\\
        \Omega_{ijk} = \frac{\mathbf{S}_i \cdot\left(\mathbf{S}_j \times \mathbf{S}_k\right)}{1 + \mathbf{S}_i \cdot \mathbf{S}_j + \mathbf{S}_j \cdot \mathbf{S}_k + \mathbf{S}_k \cdot \mathbf{S}_i}
    \end{gathered}
    \label{eq:nsk}
\end{equation}
where the summation runs over elementary triangular plaquettes in a counterclockwise direction. The topological charge signifies the existence of isolated skyrmions and other topological textures on the lattice; the skyrmion gives $n_{\rm sk}=-1$ and the antiskyrmion gives $n_{\rm sk}=1$. Due to the absence of topological textures in a purely ferromagnetic spin configuration, it has no topological charge.
The data are obtained in the ferromagnetic parameter regime, where skyrmions appear only as metastable excitations.

As a reference, it is useful to contrast our frustrated system with DM interaction-driven chiral systems. The DM interaction is defined by, 
\begin{equation}
    \mathcal{H}_\text{DM} =
    \sum_{\langle i,j\rangle}
    \mathbf{D}_{ij}\cdot
    \left(\mathbf{S}_i \times \mathbf{S}_j\right),
    \label{eq:dmi}
\end{equation}
where $\mathbf{D}_{ij}$ represents the DM vector. 
As a cross-product, the DM interaction energetically favors twisted spin textures, and the characteristic skyrmion size is typically controlled by the ratio between the DM interaction constant $D = |\mathbf{D}|$ and the ferromagnetic exchange interaction $J_1$ or simply $J$. 

In the context of chiral ferromagnets, Fujita and Sato chose LG beam widths comparable to the radius of an individual skyrmion to efficiently imprint beam profiles as skyrmionic defects and higher-order textures. In contrast, for frustrated magnets without the DM interaction, our results in Fig.~\ref{fig:is_H} indicate a qualitatively different tendency. 
Within the parameter range examined, the skyrmion generation probability increases with the beam width and is maximized for the largest value considered.

This behavior can be attributed to the fact that the skyrmion formation in the current frustrated model is heavily governed by the competing isotropic exchange interactions, $J_1$ and $J_3$.
Let's consider the spin at a conventional skyrmion center $\mathbf{S}_{\rm c}$, it's nearest neighbor $\mathbf{S}_{\rm nn}$, and third-nearest neighbor $\mathbf{S}_{\rm 3nn}$.
Due to the ferromagnetic nearest-neighbor coupling ($J_1 < 0$), it is energetically favorable to have both $\mathbf{S}_{\rm c}$ and $\mathbf{S}_{\rm nn}$ in the same direction.
On the other hand, the antiferromagnetic third-nearest-neighbor coupling ($J_3 > 0$) allows $\mathbf{S}_{\rm 3nn}$ to be anti-parallel to $\mathbf{S}_{\rm c}$.
Applying this over all the nearest neighbors and third-nearest neighbors of $\mathbf{S}_{\rm c}$, we arrive at the isolated skyrmionic defect. 
The size of this defect is given by $\sim 2a$ as inferred from the distance of the third-nearest neighbor from the skyrmion center, and observed from the final spin configuration ($t=1000$) on Fig.~\ref{fig:is_evol}.
This fixed skyrmion size is not present in chiral ferromagnets due to the lack of frustrated interactions, resulting in an optimal beam width~\cite{fujita2017}.
In our present study, the wider beams effectively induce more stochastic nucleation events, which subsequently anneals into isolated skyrmions.
Moreover, because the stabilization mechanism here does not rely on the continuous twisting favored by the small DM interaction, complex skyrmionic textures such as skyrmioniums ($2\pi$ vortices) and higher-order skyrmion multiplexes ($n\pi$ vortices) are not realized in the present frustrated system.

Another important distinction between skyrmions in chiral magnets and those in frustrated centrosymmetric systems concerns their internal degrees of freedom, such as vorticity and helicity~\cite{Okubo_PhysRevLett.108.017206}. 
In DM interaction-driven skyrmions, the sense of rotation and the helicity are essentially fixed by the form of the DM interaction~\cite{dzyaloshinskii1964theory, kataoka1981helical, bogdanov1989, bogdanov1994}, which selects a unique spin chirality and thereby locks the skyrmion texture into a specific vorticity--helicity combination. 
In contrast, skyrmions stabilized by competing symmetric interactions in frustrated magnets do not possess such intrinsic chiral preference. 
As a result, both the vorticity and helicity can, in principle, take arbitrary values, leading to a larger degeneracy of skyrmionic configurations.
Indeed, various types of skyrmions with distinct helicities and vorticities are found in Fig.~\ref{fig:is_evol}. 

Nevertheless, this degeneracy is generally lifted once additional anisotropic interactions are introduced as discussed in Sec.~\ref{sec:ja}. 
In particular, anisotropic in-plane magnetic interactions, such as bond-dependent planar anisotropy considered in Sec.~\ref{sec:ja}, can energetically favor specific orientations of the spin rotation, and thus determine the preferred vorticity and helicity in frustrated systems~\cite{Hayami_doi:10.7566/JPSJ.89.103702, Hayami_PhysRevB.105.104428}. 
Therefore, while frustrated magnets allow a broader variety of skyrmion textures compared to chiral DM systems, the actual realization of their internal structure is strongly influenced by magnetic anisotropies.

Next, we discuss the dependence of the skyrmion generation probability on the external magnetic field and the easy-axis anisotropy. 
In the ferromagnetic regime, the Zeeman coupling favors a uniformly polarized spin state, thereby suppressing topological defects such as skyrmions. 
Accordingly, as the magnetic field is increased, the system is driven deeper into the fully polarized state, and the probability of skyrmion generation is reduced, as shown in Fig.~\ref{fig:is_H}.

\begin{figure}[H]
    \centering
    \begin{minipage}{0.4\textwidth}
        \centering
        \begin{minipage}{0.45\linewidth}
            \centering
            $A=0.7, H=0.4$\\
            \includegraphics[width=\linewidth]{isolated/A=0.7/H=0.4/probability.pdf}
        \end{minipage}
        \begin{minipage}{0.45\linewidth}
            \centering
            $A=0.8, H=0.4$\\
            \includegraphics[width=\linewidth]{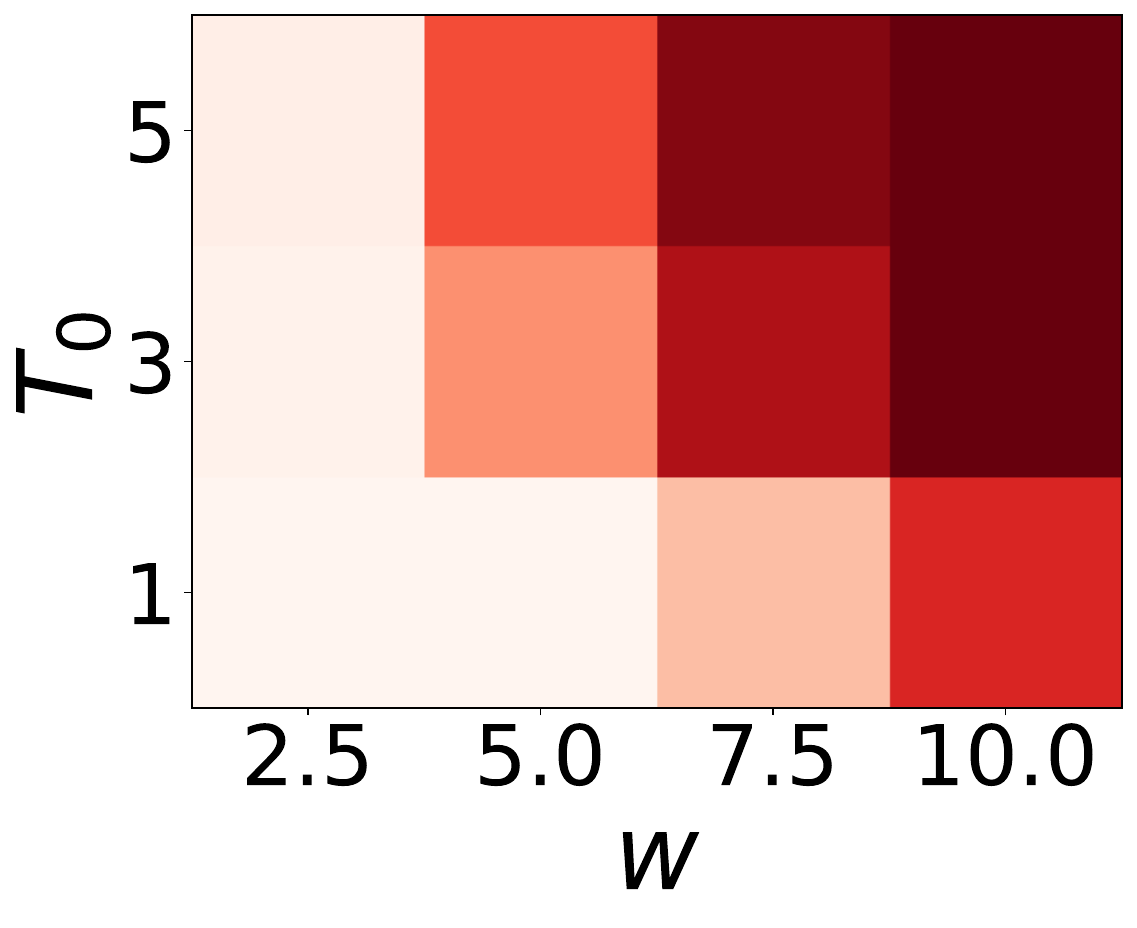}
        \end{minipage}
        \begin{minipage}{0.45\linewidth}
            \centering
            $A=0.9, H=0.4$\\
            \includegraphics[width=\linewidth]{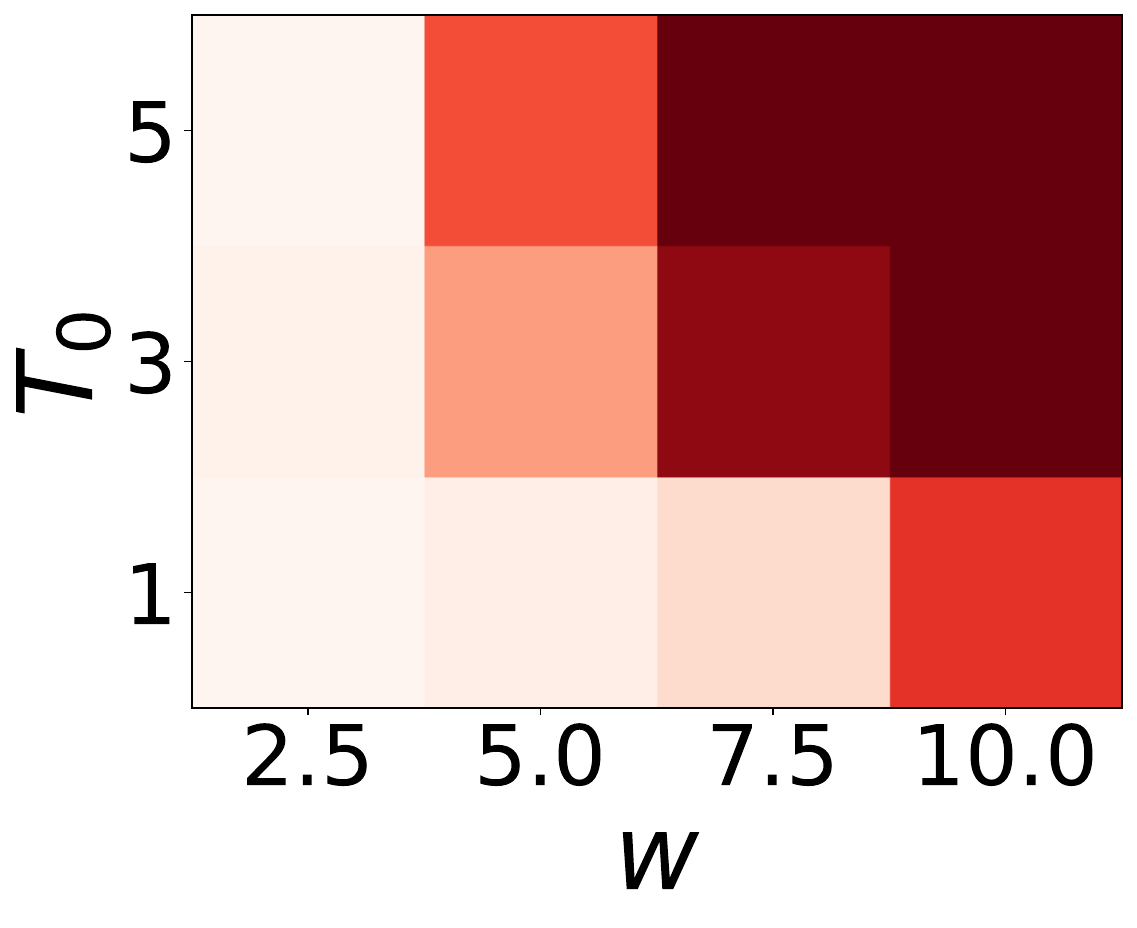}
        \end{minipage}
    \end{minipage}
    \begin{minipage}{0.075\textwidth}
        \vspace{-0.4cm}
        \includegraphics[height=2\probheight]{cbars/prob_cbar.pdf}
    \end{minipage}
    \caption{Probability maps at $H=0.3$ and $A=0.7, 0.8$, and $0.9$. The temperature is in units $J_1/k_{\mathrm{B}}$, while the beam width $w$ is in lattice unit $a$.}
    \label{fig:is_A}
\end{figure}

Figure~\ref{fig:is_A} presents the skyrmion generation probability at varying values of the easy-axis anisotropy parameter $A$. 
We find that the probability remains almost insensitive to $A$ within the examined range. 
This can be understood from the fact that easy-axis anisotropy generally supports the formation and stability of skyrmion cores by favoring out-of-plane spin components, in contrast to a strong magnetic field, which tends to eliminate anti-parallel textures.
Therefore, while increasing the field suppresses skyrmion nucleation, moderate variations in $A$ do not significantly affect the generation efficiency in the present parameter regime.

\section{Skyrmion Lattice} \label{sec:skx}

\begin{figure}[H]
    \centering
    \begin{minipage}{0.4\textwidth}
        \centering
        \begin{minipage}{0.45\linewidth}
            \centering
            $t=10$\\
            \includegraphics[width=\linewidth]{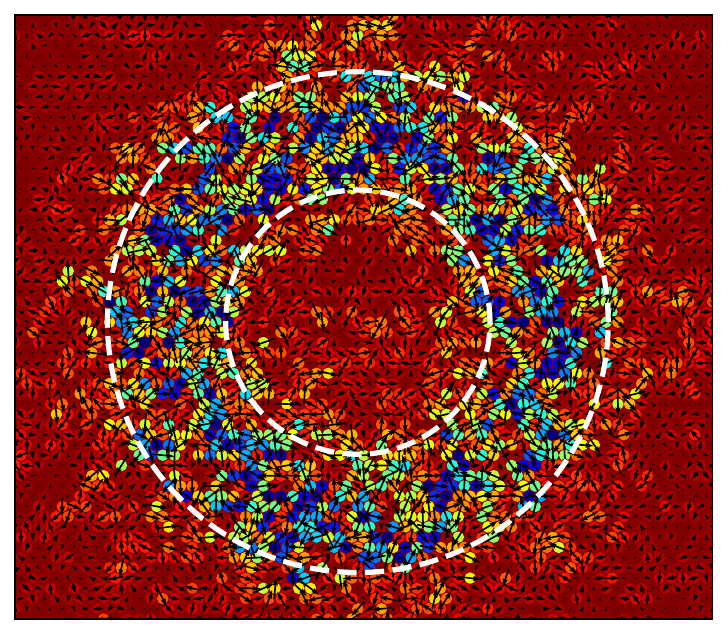}
        \end{minipage}
        \begin{minipage}{0.45\linewidth}
            \centering
            $t=200$\\
            \includegraphics[width=\linewidth]{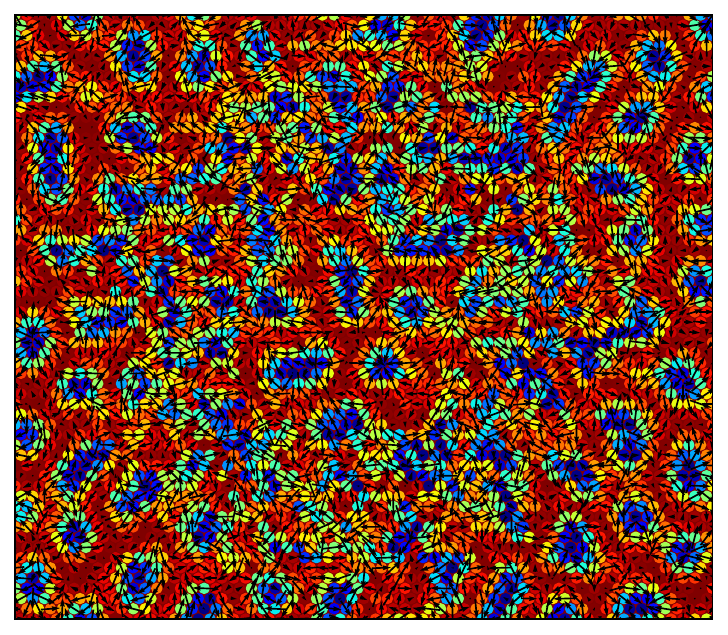}
        \end{minipage} \\
        \begin{minipage}{0.45\linewidth}
            \centering
            $t=500$\\
            \includegraphics[width=\linewidth]{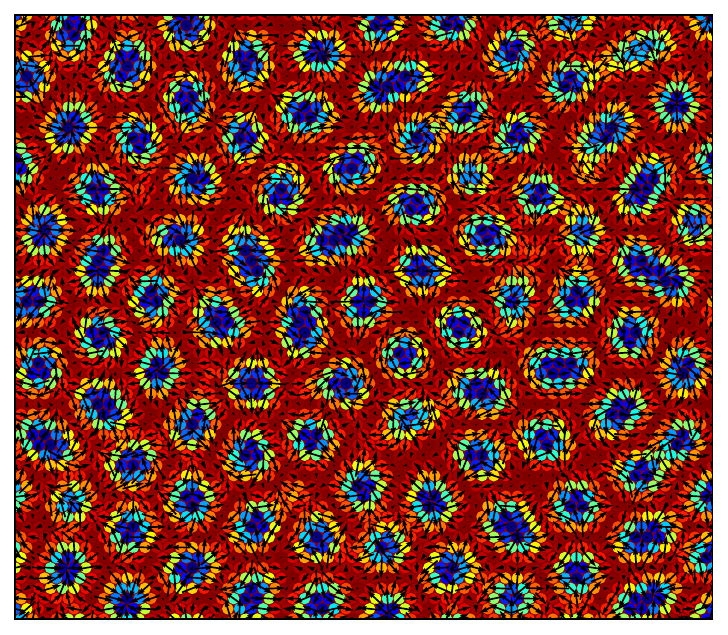}
        \end{minipage}
        \begin{minipage}{0.45\linewidth}
            \centering
            $t=1000$\\
            \includegraphics[width=\linewidth]{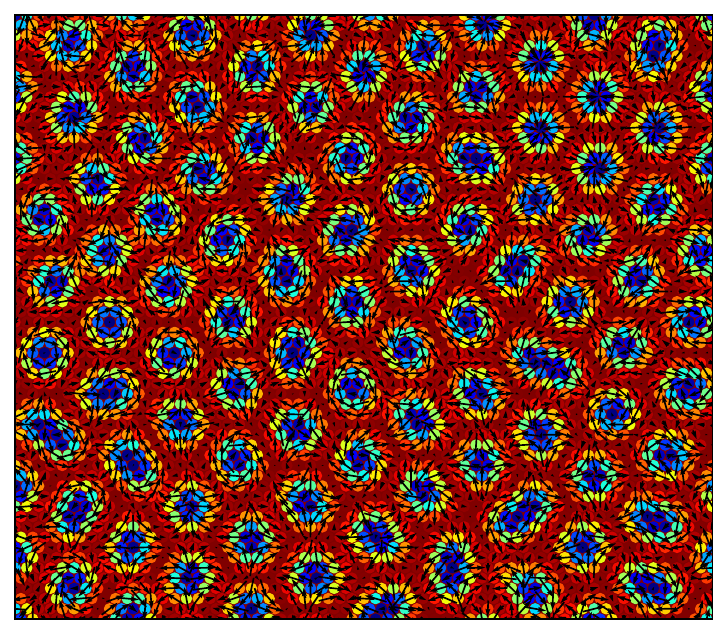}
        \end{minipage}
    \end{minipage}
    \begin{minipage}{0.075\textwidth}
        \includegraphics[height=2.2\spinheight]{cbars/spin_cbar.pdf}
    \end{minipage}
    \caption{Time evolution of a skyrmion lattice at $A=0.13$ and $H=0.35$ with beam parameters $T_0=3$, $p=0$, $m=5$, and $w=10$.}
    \label{fig:sl_evol}
\end{figure}

Figure~\ref{fig:sl_evol} shows the time evolution of a skyrmion lattice generated by LG beam irradiation in the skyrmion-lattice
regime and in its vicinity.
The simulation is performed at $A=0.13$ and $H=0.35$, where the skyrmion lattice is thermodynamically favored as the ground state.
The beam parameters are chosen as $T_0=3$, $p=0$, $m=5$, and $w=10$.
The simulations start from a ferromagnetic spin configuration, mimicking a rapid field quench from the ferromagnetic regime into the skyrmion-lattice
region.

At the initial stage of irradiation ($t=10$), the spin configuration remains close to the ferromagnetic state, although local thermal fluctuations begin to develop within the beam-irradiated region.
As the irradiation proceeds ($t=200$), nonuniform spin modulations emerge, indicating the onset of multiple-$q$ instabilities characteristic of the skyrmion-lattice phase.
Around the end of the heating process ($t=500$), skyrmion cores become clearly visible, and the system starts forming a crystalline arrangement of topological textures.
After the subsequent annealing stage ($t=1000$), the spin configuration relaxes into an ordered skyrmion lattice, representing the stable state in this parameter regime.
It should be noted that the obtained skyrmion-lattice state may form domains characterized by different vorticities, i.e, the coexistence of skyrmions and antiskyrmions.
This is because the LG beam creates skyrmions and antiskyrmions with equal probability, as shown in Sec.~\ref{sec:isolated}.

The color scale denotes the out-of-plane spin component $S^z$, which highlights the periodic array of skyrmion cores.
These results demonstrate that, in contrast to the ferromagnetic regime where isolated skyrmions are created through stochastic nucleation, skyrmion lattice 
generation occurs through a thermal annealing process that drives the system toward its equilibrium skyrmion-lattice ground state.

\begin{figure}[H]
    \centering
    \includegraphics[width=0.425\linewidth]{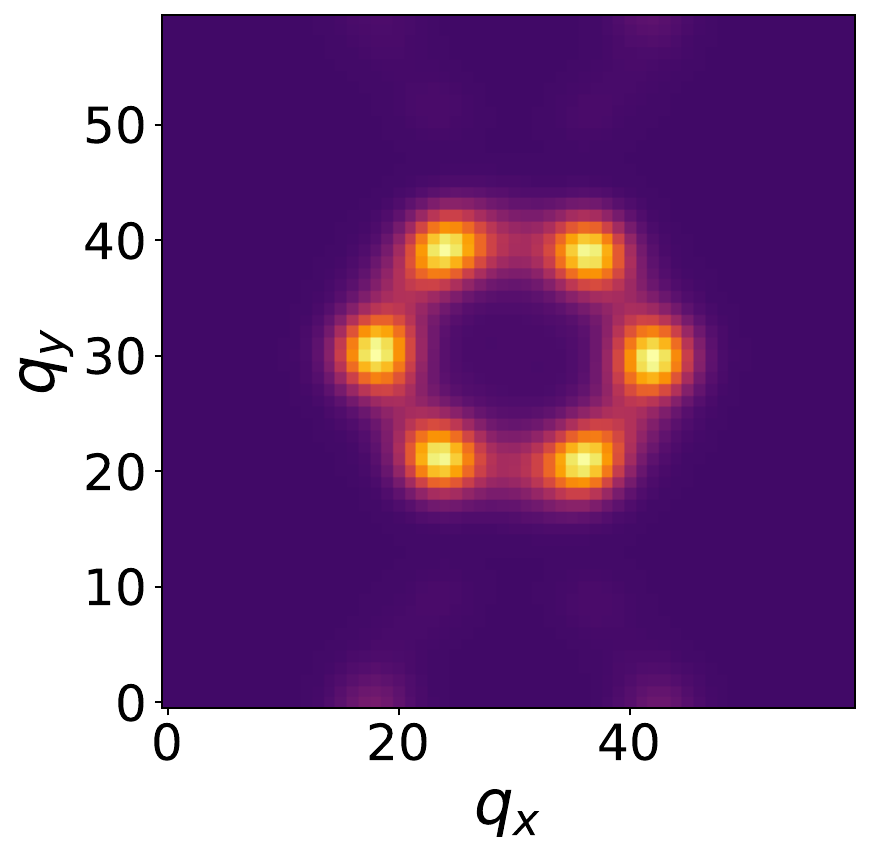}
    \includegraphics[width=0.425\linewidth]{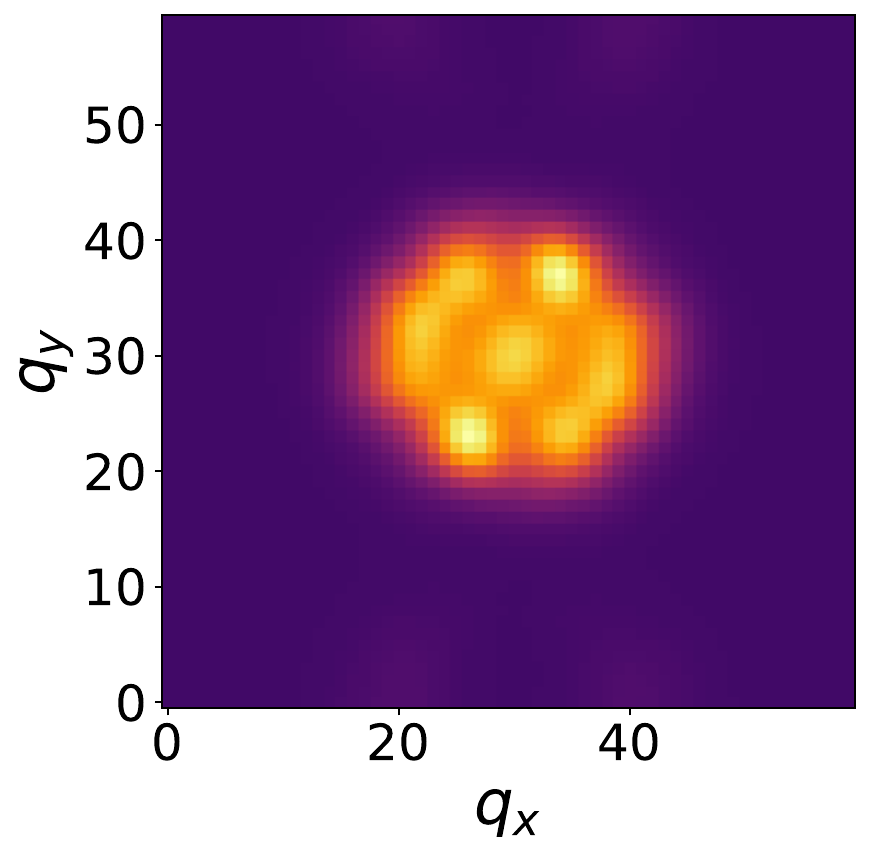}
    \includegraphics[width=0.125\linewidth]{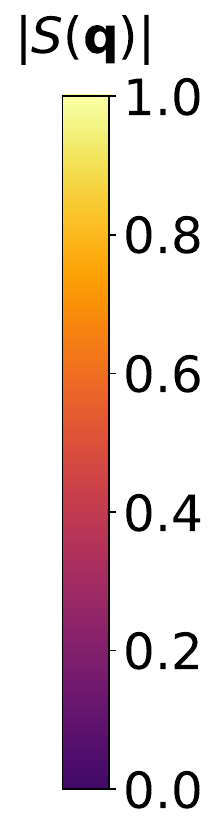}
    \caption{
    Spin structure factor $S(\mathbf{q})$ of two different simulation trials. The right figure denotes an unsuccessful trial due to the mixing of wave vector $\mathbf{q}$, whereas the left figure denotes a successful trial due to the clean separation of $6$ peaks, which is characteristic for a skyrmion lattice.
    }
    \label{fig:spin_struc}
\end{figure}

To quantify how efficiently such a skyrmion-lattice
state can be generated from the ferromagnetic spin configuration under different irradiation conditions, we next examine the success probability across various beam parameters. As discussed in Sec.~\ref{sec:isolated}, the generation probability is defined by Eq.~\eqref{eq:prob}.
However, the crystalline configuration of the skyrmion lattice requires a stricter success condition.
To determine successful trials, we evaluate the spin structure factor $S(\mathbf{q})$ defined by,
\begin{equation}
    S(\mathbf{q}) = \frac{1}{N}\left|\sum_{i}e^{i\mathbf{q}\cdot\mathbf{r}_i}\left(S_i^z-\braket{S^z}\right)\right|^2,
\end{equation}
where $\mathbf{r}_i$ denotes the distance from the origin to site $i$ and the sum is taken over every site. Figure~\ref{fig:spin_struc} shows the spin structure factor graph of a successful and an unsuccessful trial.

\begin{figure}[H]
    \centering
\begin{minipage}{0.4\textwidth}
        \centering
        \begin{minipage}{0.45\linewidth}
            \centering
            $A=0.2, H=0.3$\\
            \includegraphics[width=\linewidth]{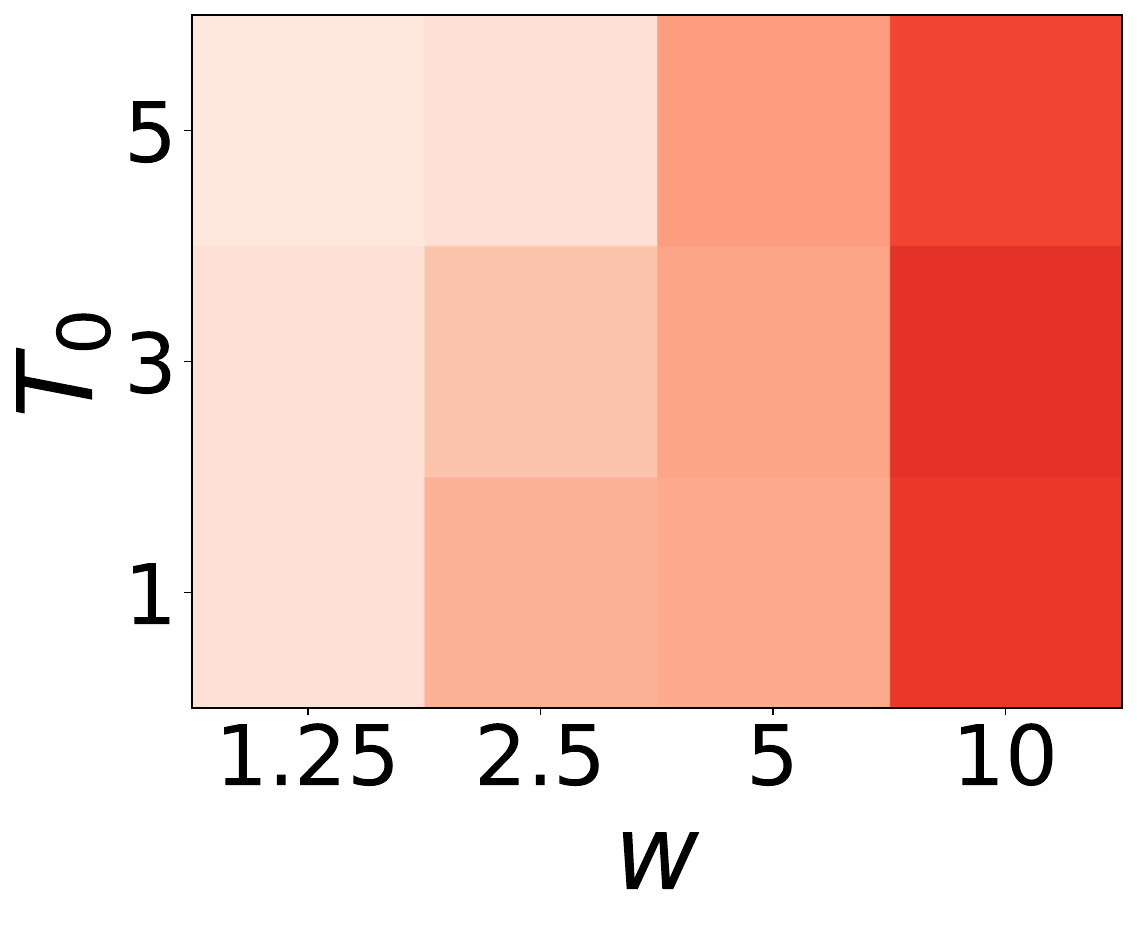}
        \end{minipage}
        \begin{minipage}{0.45\linewidth}
            \centering
            $A=0.2, H=0.4$\\
            \includegraphics[width=\linewidth]{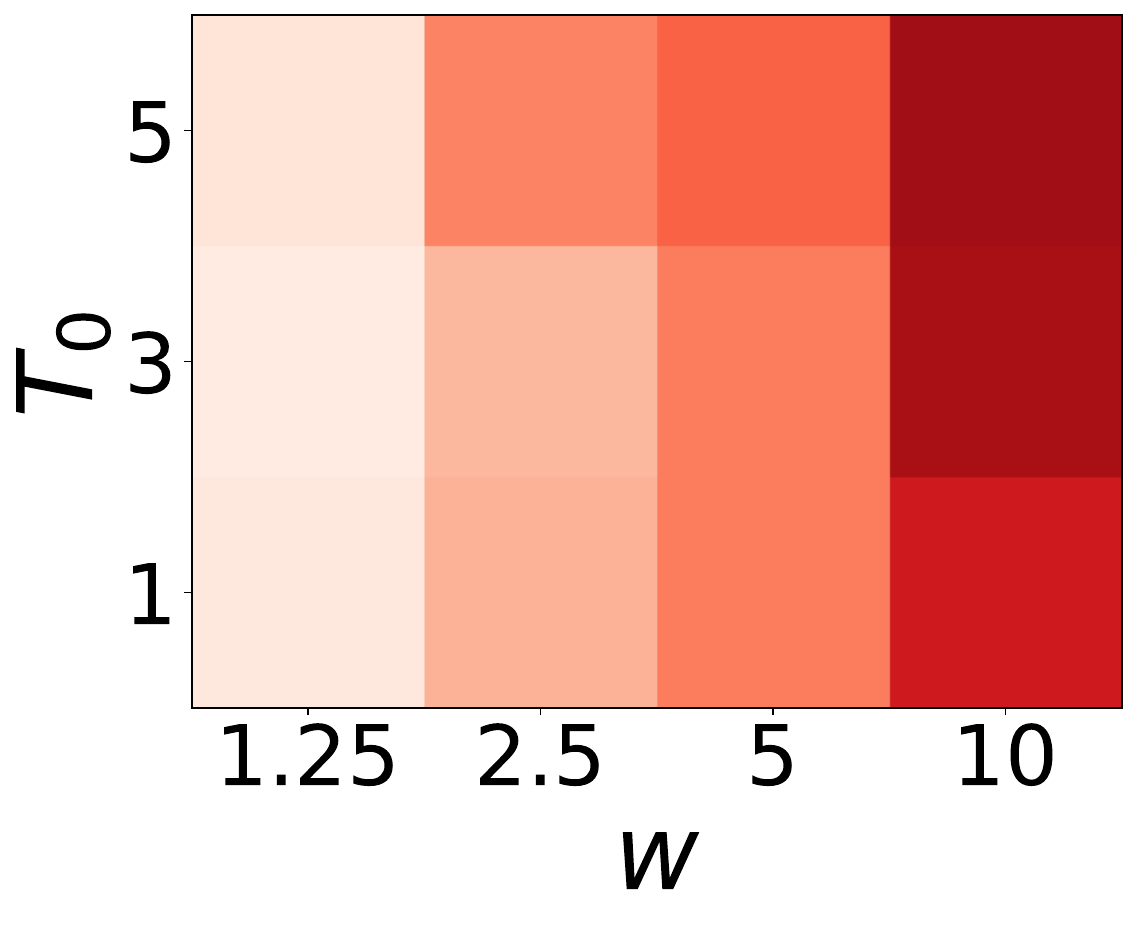}
        \end{minipage}
        \begin{minipage}{0.45\linewidth}
            \centering
            $A=0.2, H=0.5$\\
            \includegraphics[width=\linewidth]{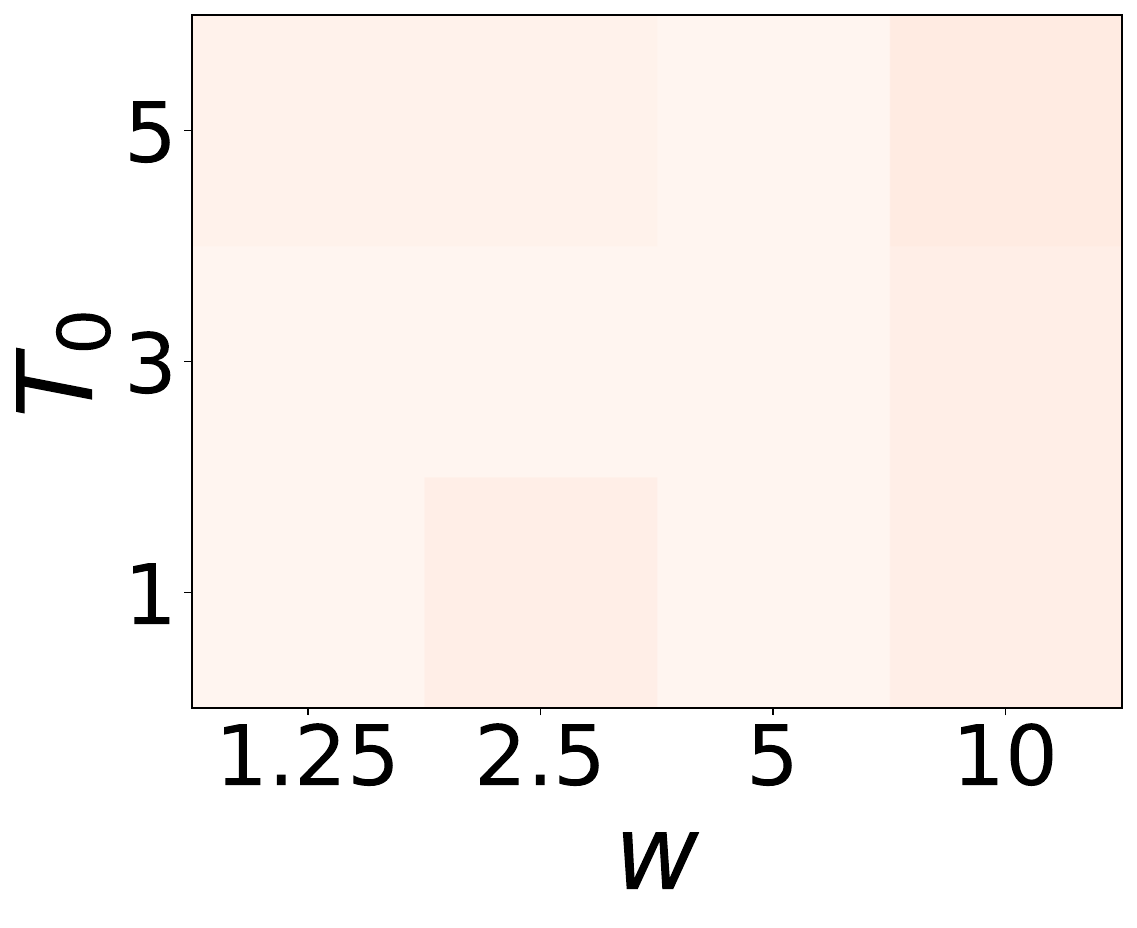}
        \end{minipage}
    \end{minipage}
    \begin{minipage}{0.075\textwidth}
        \vspace{-0.4cm}
        \includegraphics[height=2\probheight]{cbars/prob_cbar.pdf}
    \end{minipage}
    \caption{Probability maps at $H=0.3, 0.4$, and $0.5$ for $A=0.2$
    The temperature is in units $J_1/k_{\mathrm{B}}$, while the beam width $w$ is in lattice unit $a$.
    }
    \label{fig:sl_TW}
\end{figure}

The success probability is evaluated from 50 independent stochastic simulation trials for each set of beam parameters, and is defined as the percentage of runs in which an ordered skyrmion-lattice state is obtained after the irradiation and annealing processes. 
Figure~\ref{fig:sl_TW} presents the probability maps for skyrmion lattice generation under LG beam irradiation in the skyrmion-lattice regime.
These maps follow three values of the magnetic field, $H=0.3$, $0.4$, and $0.5$, a fixed easy-axis anisotropy $A=0.2$, and varying beam parameters such as $T_0 = 1, 3, 5$ and $w=1.25, 2.5, 5, 10$.
We use a non-linear scale for $w$ to showcase the non-zero probability at beam widths near the skyrmion defect size discussed in Sec.~\ref{sec:isolated}, $w\sim 2a$.
While small beam widths were not able to generate isolated skyrmions in the ferromagnetic regime, some simulations in the skyrmion-lattice regime were successful in generating skyrmion-lattice even with a small beam width.

A notable feature of Fig.~\ref{fig:sl_TW} is that the success probability exhibits a nonlinear and rather sensitive dependence on $H$.
In particular, while high magnetic fields have completely prevented skyrmion lattice annealing, low magnetic fields do not necessarily enhance the probability of skyrmion lattice formation.
This behavior contrasts with the isolated skyrmion 
case discussed in Sec.~\ref{sec:isolated}, where the generation probability linearly increased with a decrease in the applied magnetic field.
This reflects the fact that the formation of a crystalline skyrmion lattice requires a collective rearrangement of spins over the entire system, which can be hindered when the applied magnetic field excessively constrains spin fluctuations.

Furthermore, Fig.~\ref{fig:sl_TW} indicates that the skyrmion lattice generation is relatively less dependent on the detailed beam parameters such as $T_0$ and $w$, compared to the isolated skyrmion regime. 
Instead, the success of skyrmion lattice formation is primarily governed by whether the system parameters $(A,H)$ lie sufficiently close to the equilibrium skyrmion-lattice
phase.
From these probability maps, we identify $T_0=3$ and $w=10$ as optimal beam parameters for subsequent simulations exploring the broader skyrmion-lattice
region.

\begin{figure}[H]
    \centering
    \includegraphics[width=0.4\textwidth]{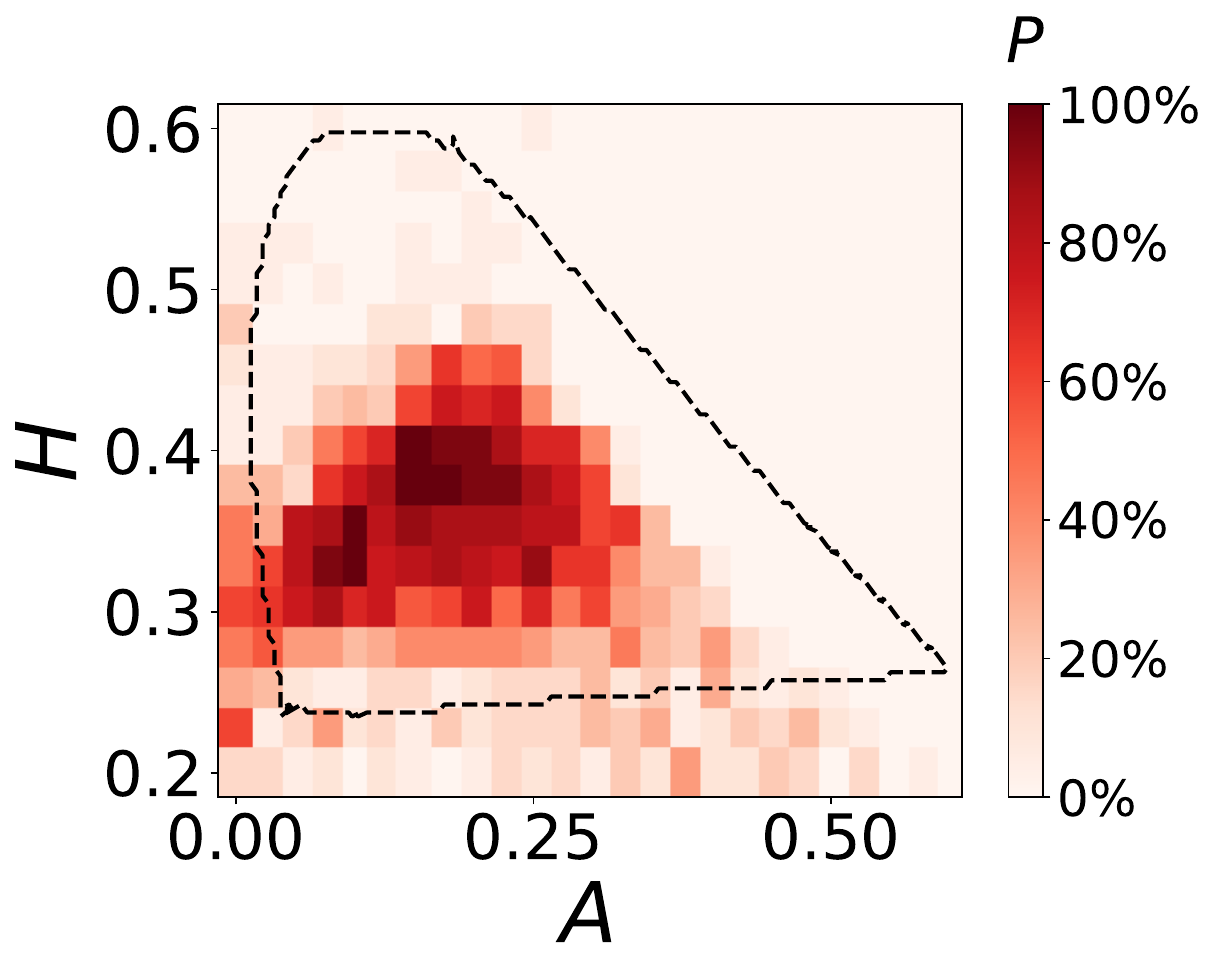}
    \caption{Probability map over $A\in[0, 0.6]$ and $H\in[0.2, 0.6]$ obtained from 20 independent trials. 
    The black region is the skyrmion lattice regime.
    }
    \label{fig:sl_AH}
\end{figure}

Having identified $T_0=3$ and $w=10$ as optimal irradiation parameters from Fig.~\ref{fig:sl_TW}, we next explore how the success probability of skyrmion lattice generation depends on the intrinsic system parameters. 
Figure~\ref{fig:sl_AH} shows the probability map in the $A$--$H$ plane, obtained from 20 independent stochastic trials at each parameter set. 
The region surrounded by the dashed black lines indicates the equilibrium skyrmion-lattice
phase in the ground-state phase diagram in Fig.~\ref{fig:phase_diagram}, providing a reference for comparison.
As seen in Fig.~\ref{fig:sl_AH}, high success probabilities are indeed concentrated within the skyrmion-lattice regime, demonstrating that LG beam irradiation can effectively drive the system from an initial ferromagnetic configuration into the topologically nontrivial skyrmion-lattice state when the parameters lie sufficiently close to the equilibrium phase. 
At the same time, the region of relatively high success probability is noticeably narrower than the skyrmion-lattice domain expected from the $T=0$ phase diagram. 
This reduction reflects the fact that, for larger easy-axis anisotropy, spins become strongly constrained along the out-of-plane direction, which suppresses the collective spin rearrangements required to establish a crystalline skyrmion texture. 
Consequently, transitioning from a uniformly polarized background into the skyrmion-lattice phase becomes increasingly difficult under strong easy-axis anisotropy, even under thermal excitation through LG beam irradiation. 

\begin{figure}[H]
    \centering
    \includegraphics[width=0.4\textwidth]{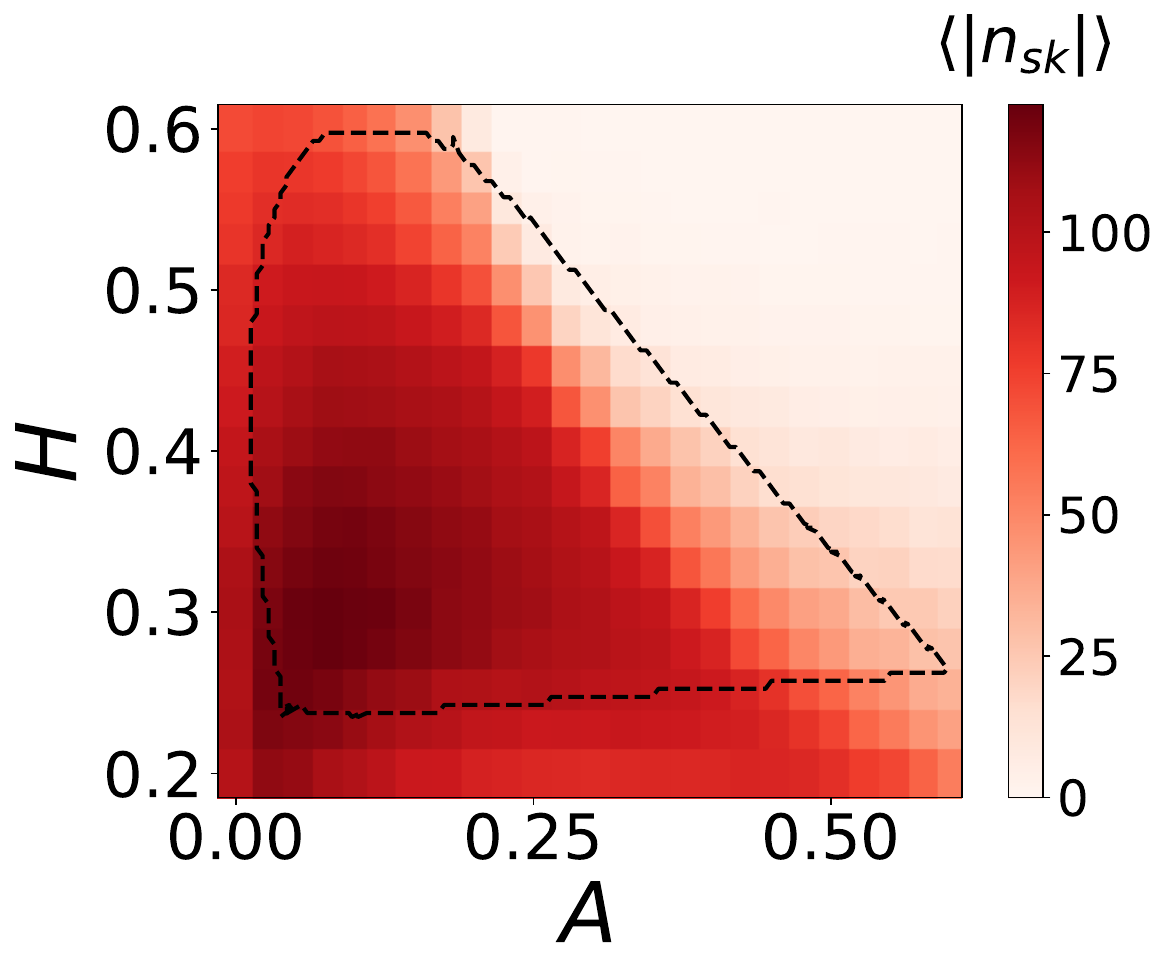}
    \caption{Field dependence of the averaged skyrmion number $\langle |n_{\rm sk}| \rangle$ and the skyrmion crystal formation probability after annealing, obtained from 20 independent trials. This figure follows the same $A$--$H$ plane as Fig.~\ref{fig:sl_AH} with $A\in[0. 0.6]$ and $H\in[0.2, 0.6]$.}
    \label{fig:sl_dens}
\end{figure}

Figure~\ref{fig:sl_dens} shows the density map of the averaged absolute skyrmion number $\langle |n_{\rm sk}| \rangle$ in the $A$--$H$ plane after LG-beam irradiation, obtained by averaging over 20 independent runs for each parameter set.
We use the absolute skyrmion number to distinguish between the ferromagnetic state ($n_{\rm sk} =0 $) and non-ferromagnetic states ($n_{\rm sk} \ne 0$), which include both skyrmions and antiskyrmions.
Furthermore, due to the $xy$-symmetry in the Eq.~\ref{eq:model}, skyrmions and antiskyrmions are degenerate.
Breaking this degeneracy will be discussed in Sec.~\ref{sec:ja}.
The color scale quantifies the absolute topological charge generated by LG beam irradiation, averaged over 20 trials.
The parameter region where $\langle |n_{\rm sk}| \rangle$ becomes large largely overlaps with the high-probability region for skyrmion lattice formation shown in Fig.~\ref{fig:sl_AH}, indicating that efficient topological-charge generation occurs near the equilibrium skyrmion-latice
regime. 

However, the two figures are not identical. 
While Fig.~\ref{fig:sl_AH} measures the probability of forming an ordered skyrmion lattice, Fig.~\ref{fig:sl_dens} captures the total amount of topological charge, regardless of spatial order. 
Consequently, even in regions where the skyrmion-lattice formation probability is low, $\langle |n_{\rm sk}| \rangle$ can remain finite, suggesting the emergence of topological textures in a more disordered form, such as coexistence states of skyrmions and helical modulations, rather than a well-developed crystal.
These results demonstrate that LG-beam irradiation can robustly create topological defects over a broad parameter range, whereas the emergence of a crystalline skyrmion latice requires more restrictive conditions.

\section{\protect Effects of Bond-dependent Planar Anisotropy} \label{sec:ja}

In this section, we discuss the effects of bond-dependent planar anisotropy on the distinction of skyrmions and antiskyrmions, and on skyrmion helicity. 
In frustrated centrosymmetric magnets, skyrmionic textures are not uniquely constrained by an intrinsic chiral interaction such as the DM coupling. 
As a consequence, their internal degrees of freedom, including vorticity and helicity, can remain highly degenerate. 
In particular, due to the $xy$-planar symmetry of the Hamiltonian in Eq.~(\ref{eq:model}), both skyrmions and antiskyrmions are energetically allowed and may appear with comparable stability in the present model.

However, from the viewpoint of discrete rotational symmetry of the lattice structure, additional bond-dependent anisotropic interactions are generally permitted in realistic materials, especially in the presence of spin--orbit coupling~\cite{kaplan1983single, Li_PhysRevB.94.035107, Michael_PhysRevB.91.155135, Yambe_PhysRevB.106.174437}. 
Such anisotropies can play an important role in selecting specific spin textures and significantly affect the stability and dynamics of skyrmions~\cite{amoroso2020spontaneous, Hayami_PhysRevB.103.054422, Utesov_PhysRevB.103.064414, Wang_PhysRevB.103.104408, yambe2021skyrmion}. 
Moreover, introducing planar anisotropy provides a natural mechanism to lift the degeneracy between skyrmions and antiskyrmions~\cite{Hayami_doi:10.7566/JPSJ.89.103702}, as well as to fix the preferred vorticity and helicity of the generated skyrmionic states.

To break the planar symmetry and incorporate these symmetry-allowed effects, we add a bond-dependent planar anisotropic term to the model in Eq.~(\ref{eq:model}), which is given by
\begin{equation}
    \mathcal{H}^{\rm BA} = J_a \sum_{\langle i,j \rangle} (\gamma_{ij}S_i^+S_j^+ + \gamma_{ij}^*S_i^-S_j^-) ,
\end{equation}
where $J_a$ is the magnitude of the bond-dependent anisotropy, $\gamma_{ij} = \gamma_{ji} = 1, e^{i2\pi/3}, e^{-i 2\pi/3}$ is the phase shift for each bond direction for any two given nearest-neighbor sites $i$ and $j$, and $S_i^{\pm} = (S_i^x \pm iS_i^y)/\sqrt{2}$.
The numerical procedure in this section follows that in Secs.~\ref{sec:isolated} and~\ref{sec:skx}. 
To focus on the role of the bond-dependent planar anisotropy without overwhelming the competing exchange interactions or the beam-induced heating effects, we restrict its magnitude to a relatively small range, $J_a \in [-0.1,0.1]$.

To distinguish skyrmions from antiskyrmions in the resulting spin configurations, we evaluate the total topological charge $n_{\rm sk}$ of the lattice as shown in Eq. \eqref{eq:nsk}. As antiskyrmions carry a topological charge opposite to skyrmions, with skyrmions having $n_{\rm sk} = -1$ and antiskyrmions having $n_{\rm sk}=1$, the average value of $n_{\rm sk}$ provides a useful quantitative measure of the relative skyrmion population. 

\begin{figure}[H]
    \centering
    \includegraphics[width=0.45\textwidth]{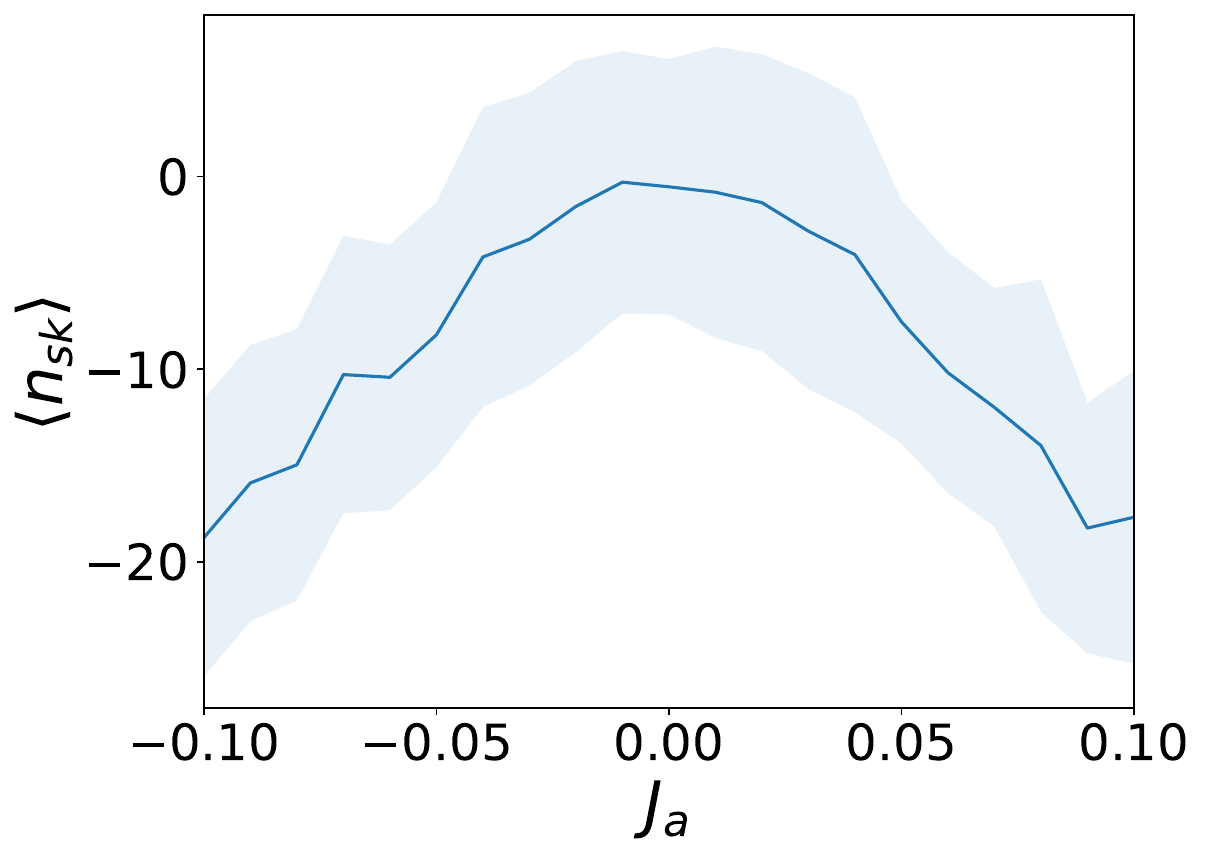}
    \caption{Planar anisotropy $J_a$ dependence of the average skyrmion number $\langle n_{\rm sk} \rangle$ across $50$ trials.
    The shaded area denotes the $n_{\rm sk}$ error per $J_a$ with 90\% confidence.}
    \label{fig:ja_before}
\end{figure}

Figure~\ref{fig:ja_before} shows the dependence of the average skyrmion number $\langle n_{\rm sk}\rangle$ on the planar anisotropy strength $J_a$, obtained from 50 independent stochastic trials for each parameter set. 
The shaded region represents the statistical uncertainty with a 90\% confidence interval.
Notably, at $J_a=0$, skyrmions and antiskyrmions are nearly degenerate due to the $xy$-planar symmetry of the Hamiltonian, and they are generated with comparable probability. 
As a result, their opposite topological charges tend to cancel out, leading to an average value $\langle n_{\rm sk}\rangle \simeq 0$.

Once a finite bond-dependent planar anisotropy is introduced, this degeneracy is lifted; the skyrmion with negative $n_{\rm sk}$ is favored compared to the antiskyrmion with positive $n_{\rm sk}$, irrespective of the sign of $J_a$. 
An increase in skyrmion population, signaled by the decrease in $\langle n_{\rm sk}\rangle$, is observed for $J_a\neq 0$.
This indicates that the anisotropy suppresses antiskyrmion formation and favors skyrmions with a particular sense of spin rotation. 
Furthermore, Fig.~\ref{fig:ja_before} demonstrates that even weak symmetry-allowed planar anisotropy ($|J_a|\ll 1$) can improve the selectivity of skyrmion generation under optical irradiation by resolving the near-degeneracy between skyrmions and antiskyrmions in frustrated magnets.

We further investigate a quenching-like situation, where the bond-dependent planar anisotropy is not present during the LG-beam irradiation process, but is instead switched on only after the irradiation has finished.
This setup mimics the case in which the planar anisotropy term is effectively induced or activated during the subsequent relaxation stage, rather than directly influencing the stochastic nucleation dynamics under the beam.

\begin{figure}[H]
    \centering
    \begin{minipage}{0.4\textwidth}
        \centering
        \begin{minipage}{0.45\linewidth}
            \centering
            $J_a = -0.05$\\
            \includegraphics[width=\linewidth]{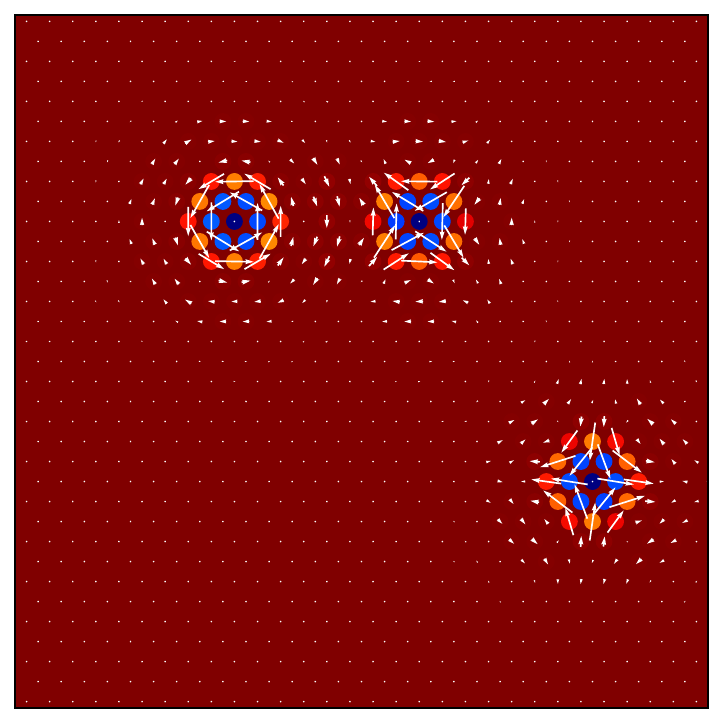}
        \end{minipage}
        \begin{minipage}{0.45\linewidth}
            \centering
            $J_a = 0.05$\\
            \includegraphics[width=\linewidth]{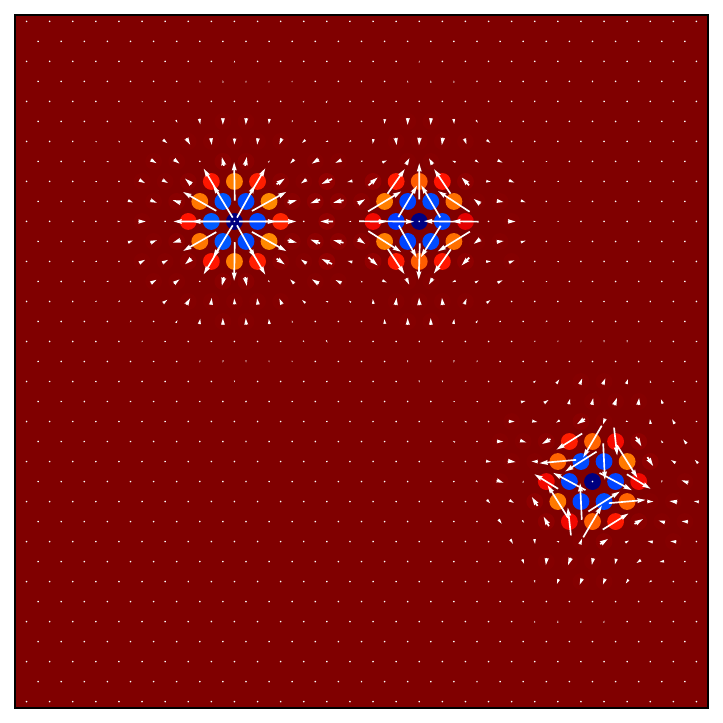}
        \end{minipage}
        \begin{minipage}{0.45\linewidth}
            \centering
            $J_a = 0$\\
            \includegraphics[width=\linewidth]{ja_test/ja_0.pdf}
        \end{minipage}
    \end{minipage}%
    \begin{minipage}{0.075\textwidth}
        \includegraphics[height=1.75\jaheight]{cbars/spin_cbar.pdf}
    \end{minipage}
    \caption{Simulations wherein planar anisotropy $J_a$ was applied only at annealing and not during beam irradiation.
    The bottom graph shows the spin configuration before annealing ($J_a = 0$).
    When annealing at $J_a = -0.05$ (upper left), the system relaxes into Bloch-type skyrmions.
    Meanwhile, annealing at $J_a = 0.05$ (upper right) results into Neel-type skyrmions.}
    \label{fig:ja_after}
\end{figure}

Figure~\ref{fig:ja_after} shows representative spin configurations obtained in such setting. 
The bottom panel corresponds to the state immediately after irradiation, where $J_a=0$ and both skyrmions and antiskyrmions can still appear due to their degeneracy.
When the system is subsequently annealed with a finite planar anisotropy, however, the resulting skyrmions exhibit a clear helicity selection.
In particular, for $J_a<0$, the skyrmions relax into a Bloch-type configuration with helicity $\gamma=\pm\pi/2$, whereas for $J_a>0$, a N\'eel-type texture with $\gamma=0$ or $\pi$ becomes energetically preferred.

Notably, unlike the case where $J_a$ is applied throughout the entire irradiation process (Fig.~\ref{fig:ja_before}), quenching does not significantly reduce the number of antiskyrmions.
This indicates that the suppression of antiskyrmion formation requires the planar anisotropy to act already at the nucleation stage.
On the other hand, once a skyrmion core is formed, the planar anisotropy can still play a crucial role in fixing its internal degree of freedom, namely the helicity, during the relaxation process.
Thus, bond-dependent planar anisotropy can serve as a possible control parameter not only for lifting the skyrmion--antiskyrmion degeneracy, but also for selecting the rotational character of optically generated skyrmions.

\section{Conclusions} \label{sec:conclusion}

We investigated the generation of skyrmionic textures in a frustrated $J_1$--$J_3$ ferromangetic system with easy-axis anisotropy through LG beam irradiation. 
The LG beam irradiation was modeled as stochastic and spatially non-uniform thermal field, and the ensuing spin dynamics were analyzed by solving the sLLG equation. 
We demonstrated that LG beam irradiation can generate two distinct classes of skyrmionic states depending on the underlying magnetic regime: isolated skyrmions in the high-field ferromagnetic phase and skyrmion lattices in the intermediate-field region where the skyrmion lattice is thermodynamically favored.

Our results highlight qualitative differences between skyrmions in frustrated magnets and DM interaction-driven skyrmions in chiral systems. 
In chiral magnets, the DM interaction fixes a sense of spin twisting, enabling efficient optical imprinting when the beam width matches the skyrmion radius~\cite{fujita2017}. 
In contrast, in frustrated centrosymmetric magnets, skyrmions are stabilized by competing symmetric exchanges and their vorticity and helicity are less constrained. 
In the ferromagnetic regime, we find that isolated skyrmions are generated via stochastic nucleation: beam-induced thermal fluctuations overcome the nucleation barrier and the defect is stabilized during relaxation. 
The probability maps show that the creation probability decreases with increasing magnetic field, while its dependence on the easy-axis anisotropy is weak in the range studied.

We also explored skyrmion-lattice generation by initializing the system in a ferromagnetic configuration, mimicking a rapid quench from the ferromagnetic regime into the skyrmion-lattice region.
In this case, LG beam irradiation effectively acts as a thermal annealing process that drives the system toward the equilibrium skyrmion lattice.
The probability maps show that the success of skyrmion-lattice formation is primarily governed by the intrinsic parameters $(A,H)$ and by the proximity to the equilibrium skyrmion-lattice phase, whereas the dependence on the beam parameters $(T_0,w)$ is relatively weaker.

Finally, we clarified the role of planar anisotropy $J_a$ in controlling the topological character of the optically generated textures. 
At $J_a=0$, the $xy$-planar symmetry renders skyrmions and antiskyrmions nearly degenerate, leading to $\langle n_{\rm sk} \rangle\simeq 0$ due to cancellation between opposite topological charges. 
Introducing a finite bond-dependent planar anisotropy lifts this degeneracy and suppresses antiskyrmion formation during nucleation, thereby enhancing selective skyrmion generation. 
Moreover, considering planar anisotropy quenching allows for helicity control even after skyrmions have already nucleated: Bloch-type skyrmions are stabilized for $J_a<0$ and N\'eel-type skyrmions are stabilized for $J_a>0$.
These findings establish structured optical heating as a viable route to create and control skyrmionic textures in frustrated magnets without relying on the DM interaction, and highlight planar anisotropy as an additional ingredient for tailoring the vorticity and helicity of the generated states.

This research was supported by JSPS KAKENHI Grants Numbers JP22H00101, JP22H01183, JP23H04869, JP23K03288, JP23K20827, and by JST CREST (JPMJCR23O4) and JST FOREST (JPMJFR2366). 

\bibliography{references} 

\end{document}